\documentclass[12pt,letterpaper]{article}
\pdfoutput=1
\usepackage[colorlinks=false,
linkcolor=red, 
citecolor=blue,
filecolor=red,
urlcolor=red,
linktoc=all, 
pdfstartview=FitV,
bookmarksopen=true]{hyperref}
\usepackage[left=2cm,top=1cm,right=3cm,nohead]{geometry}
\usepackage{xcolor,colortbl}
\usepackage[utf8]{inputenc} 
\usepackage{epsfig,latexsym,amsfonts,mathtools,amsthm,amssymb,amsbsy,multirow,slashed,color,
	mathrsfs,wasysym,textcomp,subfigure,wrapfig,comment,bbold,array,longtable,multirow,setspace,amsmath,url}
\usepackage{tabularx}
\usepackage{graphicx}
\usepackage{setspace}
\usepackage{subfigure}
\usepackage[bf]{caption}
\usepackage{braket}
\usepackage{comment}
\usepackage{float}
\usepackage{tikz,pgfplots}
\usetikzlibrary{mindmap,trees,shadows}
\colorlet{color1}{gray!25}
\usetikzlibrary{positioning}
\usetikzlibrary{intersections} 

\newlength{\PicScale}
\usepackage[UKenglish]{babel}
\usepackage[toc,page]{appendix}
\usepackage{hhline}
\usepackage{geometry}
\usepackage{cite}
\usepackage{datetime}
\usepackage{bbm} 
\usepackage{bm} 
\hypersetup{
	pdftitle={},
	pdfauthor={},
	pdfsubject={}
} 
\usepackage{enumitem}

\definecolor{Gray}{gray}{0.94}

\newcommand{\mra}{{\mathrm{A}}}
\newcommand{\mrd}{{\mathrm{D}}}
\newcommand{\mre}{{\mathrm{E}}}
\newcommand{\mrc}{{\mathrm{C}}}
\newcommand{\mrb}{{\mathrm{B}}}
\newcommand{\mrf}{{\mathrm{F}}}
\newcommand{\mrg}{{\mathrm{G}}}

\newcommand{\rii}{{\mathrm{II}}}

\newcommand{\uo}{{\mathrm{U}(1)}}

\newcommand{\sug}{{\mathrm{SU}}}
\newcommand{\spg}{{\mathrm{Sp}}}
\newcommand{\sog}{{\mathrm{SO}}}
\newcommand{\sping}{{\mathrm{Spin}}}

\newcommand{\tC}{{\texttt{C}}}

\newcolumntype{M}[1]{>{\centering\arraybackslash}m{#1}}
\newcolumntype{N}{@{}m{0pt}@{}}

\numberwithin{equation}{section}

\makeatletter
\def\@cline#1-#2\@nil{
	\omit
	\@multicnt#1
	\advance\@multispan\m@ne
	\ifnum\@multicnt=\@ne\@firstofone{&\omit}\fi
	\@multicnt#2
	\advance\@multicnt-#1
	\advance\@multispan\@ne
	\leaders\hrule\@height\arrayrulewidth\hfill
	\cr
	\noalign{\nobreak\vskip-\arrayrulewidth}}
\makeatother

\graphicspath{{./Images/}}

\begin{document}
\pagestyle{empty}
\begin{center}        
  {\bf\LARGE Freezing of Gauge Symmetries in the \\ Heterotic String on $T^4$\\ [3mm]}

\large{ Bernardo Fraiman$^{*\,\dagger}$ and H\'ector Parra De Freitas$^{*}$
 \\[2mm]}
{\small $*$  Institut de Physique Th\'eorique, Universit\'e Paris Saclay, CEA, CNRS\\ [-2mm]}
{\small\it  Orme des Merisiers, 91191 Gif-sur-Yvette CEDEX, France.\\[0.2cm] } 
{\small  $\dagger$ Instituto de Astronom\'ia y F\'isica del Espacio (IAFE-CONICET-UBA)\\ [-2mm]}
{\small \hspace{1em}  Departamento de F\'isica, FCEyN, Universidad de Buenos Aires (UBA) \\ [-2mm]}
{\small\it Ciudad Universitaria, Pabell\'on 1, 1428 Buenos Aires, Argentina\\ }

{\small \verb"bfraiman@iafe.uba.ar, hector.parradefreitas@ipht.fr"\\[-3mm]}
\vspace{0.3in}

\small{\bf Abstract} \\[3mm]\end{center}
We derive a map relating the gauge symmetry groups of heterotic strings on $T^4$ to other components of the moduli space with rank reduction. This generalizes the results for $T^2$ and $T^3$ which mirror the singularity freezing mechanism of K3 surfaces in F and M-theory, respectively. The novel feature in six dimensions is that the map explicitly involves the topology of the gauge groups, in particular acting only on non-simply-connected ones. This relation is equivalent to that of connected components of the moduli space of flat $G$-bundles over $T^2$ with $G$ non-simply-connected. These results are verified with a reasonably exhaustive list of gauge groups obtained with a moduli space exploration algorithm.

\newpage



\setcounter{page}{1}
\pagestyle{plain}
\renewcommand{\thefootnote}{\arabic{footnote}}
\setcounter{footnote}{0}

\tableofcontents	
\newpage

\section{Introduction}
\label{sec:intro}
The moduli space of heterotic string compactifications with maximal supersymmetry consists of different connected components which are labeled by the number of compact dimensions $d \geq 1$ and the rank $r\leq 16+d$ of the gauge group. In rare cases there are more than one component with the same values for $d$ and $r$, e.g. for $d = 7$ and $r = 3$ there are two\cite{deBoer:2001wca}. For each value of $d$ there is one component with $r = 16+d$, the so-called Narain moduli space, which is realized by compactifying on $T^d$ \cite{Narain:1985jj,Narain:1986am}. The components with reduced gauge group rank can be realized by asymmetric orbifolds\cite{Narain:1986qm} of the $T^d$ compactification involving outer automorphisms of the Narain lattice (most notably the CHL string \cite{Chaudhuri:1995fk,Chaudhuri:1995bf}), and for larger $d$ one generically finds more components with various values of $r$. 

A natural question is what gauge groups can be realized in each component of this moduli space. It has been known for a long time that in the case of $T^d$ compactifications the allowed gauge algebras correspond to root lattices which embed into the Narain lattice (see e.g. \cite{Taylor:2011wt}). For each algebra, the global data that specifies the full gauge group (i.e. its fundamental group) can similarly be obtained from the Narain lattice \cite{Font:2020rsk,Font:2021uyw}. These statements generalize to the reduced rank components for which there are analogues of the Narain lattice\cite{Mikhailov:1998si,deBoer:2001wca}, but in both cases some subtleties have to be taken into account, ultimately due to the fact that these lattices are generically not self-dual \cite{Font:2021uyw,Cvetic:2021sjm}.

For $d = 1$ the Narain lattice $\Gamma_{1,17} \simeq \Gamma_{1,1} \oplus \mre_8 \oplus \mre_8$ is Lorentzian, and its automorphism group is Coxeter, allowing to obtain all the possible gauge algebras from a Generalized Dynkin Diagram (GDD)\cite{Cachazo:2000ey}. The situation for the CHL string is yet simpler, since the momentum lattice is $\Gamma_{1,9} \simeq \Gamma_{1,1} \oplus \mre_8$\cite{Mikhailov:1998si}, and one finds in particular that all gauge groups are simply-connected (so that the algebras completely specify them)\cite{Font:2021uyw}. For $d \geq 2$ however the momentum lattices are not Lorentzian and such overarching GDDs do not exist. Instead one finds that there are many different GDDs for each moduli space component, each giving partial information \cite{Font:2020rsk,Font:2021uyw}. Still, however, these diagrams do not yield the full list of gauge groups in a controlled way. 

In order to obtain the full list of gauge algebras, an algorithm was developed in \cite{Font:2020rsk,Font:2021uyw,Fraiman:2021soq}, which basically works by taking a point in the moduli space with maximally enhanced gauge symmetry, i.e. with no $\uo$ factors, and looking for others in its vicinity. By iterating this algorithm one is able to obtain all such maximal enhancements in $d = 1,2,3$ for the Narain and CHL components \cite{Font:2021uyw,Font:2020rsk,Fraiman:2021soq}. In \cite{Fraiman:2021soq} we obtained the full gauge groups for all the moduli space components in $d = 3$, basing our methods on the results of \cite{Cvetic:2021sjm} for those with rank reduction.

It turns out that for $d = 1,2,3$ all the gauge groups of the reduced rank components can be obtained from those of the Narain component by means of a suitable map. At the level of the algebras this has been known for a long time, for $d = 2$, in the dual frame of F-theory on K3, where reduced rank algebras are obtained by partially ``freezing" the singular fibers \cite{Witten:1997bs,Bhardwaj:2018jgp}. This situation extends to $d = 3$ using M-theory on K3\cite{Atiyah:2001qf,deBoer:2001wca,Tachikawa:2015wka}. In the heterotic string these results can be reproduced by using lattice embedding techniques, and in fact one can also see how the full gauge group is ``frozen". For $d = 2$ this extension was done in \cite{Cvetic:2021sjm}, and we have found similar results for $d = 3$ in \cite{Fraiman:2021soq}. As explained in the text, for $d = 1$ it suffices to delete an $\mre_8$ factor in any gauge group containing it, and both groups related by the map are simply-connected. 

The purpose of this work is to extend these results to $d = 4$, focusing on the map from the Narain component to the rank reduced components of the moduli space. The main novelty in $d = 4$ is that in order to apply the map one must know the fundamental group of the gauge group explicitly. It cannot be naively applied at the level of the algebras as in $d = 1,2,3$. This is due to the fact that the lattice which corresponds to the frozen singularity in the heterotic frame is a root lattice for $d = 1,2,3$ but the weight lattice of a non-simply-connected group for $d = 4$. Most remarkably, however, is the fact that the maps seem to be exactly those which relate the ``topologically nontrivial" components of the moduli space of flat connections of a non-simply-connected gauge group over $T^2$ (and not $T^4$) to the ``topologically trivial" component\cite{Schweigert:1996tg,Lerche:1997rr}, although to our awareness an explanation for this is lacking.

For this work we have carried out an exploration of the possible maximally enhanced gauge groups realized in six components of the moduli space of heterotic strings with 16 supercharges, which can be accessed at \cite{fp2021}. Although we cannot guarantee its exhaustivity, it is extensive enough to check the freezing map we have derived.

This paper is structured as follows. In Section \ref{sec:CHL} we explain how the freezing map is constructed for generic $d$, reviewing the cases $d = 2,3$ and deriving the case $d = 4$. We also give the map for $d = 1$. In Section \ref{sec:others} we extend the freezing map to the 6d heterotic vacua obtained by compactifying the 7d holonomy triples of \cite{deBoer:2001wca} to 6d. Our results are summarized and discussed in Section \ref{sec:results}.

\newpage
\section{Mapping gauge groups from Narain to CHL}
\label{sec:CHL}
In this section we explain the general method for determining the map which connects the Narain component with the CHL component and explicitly derive it for $d = 1,2,3,4$. The case $d = 2$ was first obtained in \cite{Cvetic:2021sjm} and the case $d = 3$ in \cite{Fraiman:2021soq}. Extensions to other rank reduced components are considered in Section \ref{sec:others}.

\subsection{Setup and basic facts}

In order to determine the map which applies to the gauge groups of the Narain component of the moduli space to give those of the rank reduced components we have to relate the way in which these are obtained in each case from the corresponding momentum lattices. We will illustrate this procedure using the CHL string, and so the focus is on the Narain lattice $\Gamma_N$ and the Mikhailov lattice $\Gamma_M$, which can be written as
\begin{equation}\label{NMlattices}
	\begin{split}
		\Gamma_N &\simeq \Gamma_{d,d}\oplus \mre_8 \oplus \mre_8 \,,\\
		\Gamma_M &\simeq \Gamma_{d-1,d-1}(2) \oplus \Gamma_{1,1} \oplus \mre_8\,.
	\end{split}
\end{equation}
Here $\Gamma_{d,d}\simeq \bigoplus_{i = 1}^d \Gamma_{1,1}$ is the unique even self-dual lattice with signature $(+^d,-^d)$, where $\Gamma_{1,1}$ is the hyperbolic lattice with Gram matrix $\begin{psmallmatrix}0&1\\1&0\end{psmallmatrix}$. The symbol $(2)$ denotes a rescaling of the lattice by $\sqrt{2}$, hence a rescaling of the Gram matrix by $2$. The lattice $\mre_8$ is just the lattice generated by the roots of the algebra $\mathfrak{e}_8$, but for the latter, as well as for the groups, we will use the symbol $\mre_8$ when there is no risk of ambiguity. The same applies for any other root lattice of A-to-G type. We convene in taking the momentum lattices to have signature with mostly pluses, unless stated otherwise.

For the Narain component of the moduli space one obtains all the possible gauge algebras by finding embeddings of root lattices $L$ into $\Gamma_N$ such that the intersection of $L \otimes \mathbb{R}$ with $\Gamma_N$ is an overlattice $M \hookleftarrow L$ whose maximal root sublattice is $L$ itself. Here we mean by overlattice any lattice of the same rank containing the lattice in question. Intersections of real slices such as $L \otimes \mathbb{R}$ with $\Gamma_N$ give lattices which are said to be primitively embedded, in this case in $\Gamma_N$, hence the embedding $M \hookrightarrow \Gamma_N$ is primitive but $L \hookrightarrow \Gamma_N$ is not unless $M = L$. By roots we mean vectors $v \in \Gamma_N$ with norm $v\cdot v = 2$, since these are the ones associated to root states in the adjoint representation of the gauge algebra.

This discussion extends to the CHL component of the moduli space, with the only difference being that roots are not only vectors with norm 2 but also vectors $v$ with norm 4 satisfying the condition $v \cdot u = 0 \mod 2$ for all vectors $u \in \Gamma_M$ \cite{Mikhailov:1998si}. This last condition is equivalent to the statement that the coroot $v^\vee = \tfrac12 v$ is in the dual lattice $\Gamma_M^*$, which is the language used in \cite{Cvetic:2021sjm}. Note that $v^\vee \cdot v^\vee = 1$, hence this condition cannot be satisfied by any vector in the Narain lattice which is even and self-dual. The same applies to $\Gamma_M$ when $d = 1$. For $d \geq 2$, however, $\Gamma_M$ is not self dual and $\Gamma_M^*$ indeed contains vectors with norm 1. The appearance of non-simply-laced algebras seems therefore to be intimately connected with the non-self-duality of the momentum lattice for the moduli space component in question. 

These facts allow to obtain the possible gauge algebras $\mathfrak g$ in these moduli space components, but we are also interested in the full gauge groups $G$. For this we need to compute the fundamental group $\pi_1(G)$, which we denote by $H$. If $\tilde G$ is the universal cover of $G$, then $G = \tilde G/H$. In the Narain component it suffices to compute the lattice quotient $M/L$, which gives a finite Abelian group isomorphic to $H$ due to the self-duality of $\Gamma_N$ \cite{Font:2021uyw}. For example, if $M = L$, then $G$ is simply-connected. For the CHL component one must do a more precise analysis \cite{Cvetic:2021sjm}, but the upshot is that $H$ is given by the quotient $M^\vee / L^\vee$, where $L^\vee$ is the coroot lattice of $\mathfrak g$ embedded in the dual lattice $\Gamma_M^*$, and $M^\vee$ its overlattice which embeds primitively into $\Gamma_M^*$. Clearly, this is a generalization of the computation for $\Gamma_N$. In both cases $H$ is a subgroup of the center $Z(G)$, specified by a set of elements $k_i \in Z(G)$.

\subsection{Construction of the map in \texorpdfstring{$d = 1,2,3$}{d=1,2,3}}

To relate the Narain and the CHL constructions just outlined we require some additional facts. For $d = 1, 2, 3, 4$, $\Gamma_M$ can be written respectively as \cite{Mikhailov:1998si}
\begin{equation}
	\Gamma_M \simeq \Gamma_{d,d}\oplus \Lambda\,,~~~~~ ~~~~~ \Lambda = 
	\begin{cases}
		\mre_8 & ~~~~~ d = 1\\
		\mrd_8 & ~~~~~ d = 2\\
		\mrd_4 \oplus \mrd_4  & ~~~~~ d = 3\\
		\mrd_8^*(2) & ~~~~~ d = 4
	\end{cases}\,,
\end{equation}
to which we restrict our attention in the following. In each case there is an embedding 
\begin{equation}\label{singemb}
	\Gamma_M \oplus \Lambda \hookrightarrow \Gamma_N\,,
\end{equation} 
where $\Gamma_M \hookrightarrow \Gamma_N$ and $\Lambda \hookrightarrow \Gamma_N$ are primitive. Furthermore, the primitive embedding of $\Lambda$ into $\Gamma_N$ is unique (up to automorphisms of $\Gamma_N$), so that by constructing any such embedding one may take its orthogonal complement which by necessity is just $\Gamma_M$. As we will review, $\Lambda$ can be interpreted as the K3 frozen singularitiy (or singularities) in the dual geometric frame both for 8d and 7d, and so we will refer to it as the \textit{frozen sublattice} in the heterotic string context. We also use the terms mapping (from Narain to reduced rank) and freezing interchangeably.

Consider now a lattice\footnote{Here we prime the lattice $M$ in the Mikhailov lattice since we will later focus on the map \textit{to} and not \textit{from} the CHL component.} $M'$ primitively embedded into $\Gamma_M$, with root sublattice $L'$. It follows from \eqref{singemb} that there is an embedding
\begin{equation}
	M' \oplus \Lambda \hookrightarrow \Gamma_N
\end{equation}
with $M'$ (but not necessarily $M'\oplus \Lambda$) primitively embedded into $\Gamma_N$. The intersection $(M'\oplus \Lambda)\otimes \mathbb{R}\cap \Gamma_N$ gives a lattice $M$ primitively embedded into $\Gamma_N$, with root sublattice $L$. This gives a priori a map $\varphi$ from a gauge algebra $\mathfrak g_\text{CHL}$ in CHL moduli space to another $\mathfrak{g}_\text{Narain}$ in Narain moduli space, but since we are dealing with the full embedding data for each lattice, we can also obtain the fundamental group of the gauge group and promote this map to one at the level of groups,
\begin{equation}
	\varphi: ~~~ G_\text{CHL} \mapsto G_\text{Narain}\,.
\end{equation} 

Consider conversely a lattice $M$ primitively embedded into $\Gamma_N$, with root sublattice $L$, such that $\Lambda$ is in turn primitively embedded into $M$ (note that primitivity in this case is guaranteed by the fact that $\Lambda \hookrightarrow \Gamma_N$ is primitive). It follows that $M$ has a sublattice of the form $M'\oplus \Lambda$, where both $M'$ and $\Lambda$ are primitively embedded into $M$. Since the orthogonal complement of $\Lambda$ in $\Gamma_N$ is just $\Gamma_M$, it follows that $M'$ is primitively embedded into $\Gamma_M$, and defines a gauge group $G_\text{CHL}$. This gives a map
\begin{equation}
	\varphi^{-1}: ~~~ G_\text{Narain} \mapsto G_\text{CHL}\,.
\end{equation}
We note however that the embedding $\Lambda \hookrightarrow M$ is not necessarily unique so that this map is generically one-to-many. As we will see, the form of this map has markedly different qualitative features depending on the value of $d$. In the following we study explicitly the cases $d = 1,2,3,4$. 

\subsubsection{$d = 1$}

For $d = 1$, we have that $\Gamma_M \simeq \Gamma_{1,1} \oplus \mre_8$ and $\Lambda = \mre_8$ are even self-dual. Therefore, eq. \eqref{singemb} can be replaced by a stronger statement (cf. eq. \eqref{NMlattices}),
\begin{equation}\label{NMd1}
	\Gamma_N \simeq \Gamma_M \oplus \mre_8\,, ~~~~~ d = 1\,.
\end{equation}
In this case, the lattice $M'$ that we consider is a root lattice $L'$, since in nine dimensions all gauge groups are simply-connected \cite{Font:2021uyw}. Therefore we have an embedding $L'\oplus \mre_8 \hookrightarrow \Gamma_N$. This embedding \textit{is} primitive, since $L'\hookrightarrow\Gamma_M$ is primitive and $\mre_8$ is unimodular, so there does not exist an even overlattice of $L'\oplus \mre_8$ in $\Gamma_N$. Moreover, $L'\oplus \mre_8$ is a root lattice corresponding to a simply-connected gauge group in Narain moduli space. We see therefore that to every gauge group $G_\text{CHL}$ in the CHL component we can associate another group  $G_\text{Narain}$ in the Narain component by some map
\begin{equation}
	\varphi: ~~~ G_\text{CHL} \mapsto G_\text{Narain} = G_\text{CHL} \times \mre_8\,, ~~~~~d = 1.
\end{equation}
Conversely, consider some root lattice of the form $L'\oplus \mre_8$ in $\Gamma_N$. Similarly to the CHL component, all of the associated groups are simply-connected. Since the primitive embedding of $\mre_8$ into $\Gamma_N$ is unique, it follows that $L'$ is primitively embedded into $\mre_8^\perp \simeq \Gamma_M$. This means that any gauge group of the form $G\times \mre_8$ in the Narain component necessarily has $G = G_\text{CHL}$ some group in the CHL component. At the end of the day, the result is that by taking all gauge groups in the Narain component which contain an $\mre_8$ factor and deleting it one obtains all of the gauge groups in the CHL component. If there are two $\mre_8$ factors, they are equivalent by an automorphism of $\Gamma_N$, so that there is no ambiguity in deleting one or the other. 

\begin{figure}
	\centering
	\begin{tikzpicture}[scale = 1.25]
		\draw(0,0)--(7,0);
		\draw(0.5,0)--(0.5,1);
		\draw(6.5,0)--(6.5,0.5);
		\draw(6.5,0.5)--(6.5,1);
		\draw[fill=white](0,0) circle (0.1) node[below = 0.25]{\small{1}};
		\draw[fill=white](0.5,0) circle (0.1) node[below = 0.25]{\small{2}};
		\draw[fill=white](1,0) circle (0.1) node[below = 0.25]{\small{3}};
		\draw[fill=white](1.5,0) circle (0.1) node[below = 0.25]{\small{4}};
		\draw[fill=white](2,0) circle (0.1) node[below = 0.25]{\small{5}};
		\draw[fill=white](2.5,0) circle (0.1) node[below = 0.25]{\small{$6$}};
		\draw[fill=white](3,0) circle (0.1) node[below = 0.25]{\small{$0$}};
		\draw[fill=white](3.5,0) circle (0.1) node[below = 0.25]{\small{\tC}};
		\draw[fill=white](4,0) circle (0.1) node[below = 0.25]{\small{$0'$}};
		\draw[fill=white](4.5,0) circle (0.1) node[below = 0.25]{\small{$6'$}};
		\draw[fill=white](5,0) circle (0.1) node[below = 0.25]{\small{$5'$}};
		\draw[fill=white](5.5,0) circle (0.1) node[below = 0.25]{\small{$4'$}};
		\draw[fill=white](6,0) circle (0.1) node[below = 0.25]{\small{$3'$}};
		\draw[fill=white](6.5,0) circle (0.1) node[below = 0.25]{\small{$2'$}};
		\draw[fill=white](7,0) circle (0.1) node[below = 0.25]{\small{$1'$}};
		\draw[fill=white](6.5,0.5) circle (0.1) node[right = 0.25]{\small{$7'$}};
		\draw[fill=white](6.5,1) circle (0.1) node[right = 0.25]{\small{$8'$}};
		\draw[fill=white](0.5,0.5) circle (0.1) node[left = 0.25]{\small{$7$}};
		\draw[fill=white](0.5,1) circle (0.1) node[left = 0.25]{\small{$8$}};
		\draw(3.5,1)node{$\Gamma_N$};
		%
		%
		\begin{scope}[shift={(9,0)}]
			\draw(0,0)--(3.5,0);
			\draw(0.5,0)--(0.5,1);
			\draw[fill=white](0,0) circle (0.1) node[below = 0.25]{\small{1}};
			\draw[fill=white](0.5,0) circle (0.1) node[below = 0.25]{\small{2}};
			\draw[fill=white](1,0) circle (0.1) node[below = 0.25]{\small{3}};
			\draw[fill=white](1.5,0) circle (0.1) node[below = 0.25]{\small{4}};
			\draw[fill=white](2,0) circle (0.1) node[below = 0.25]{\small{5}};
			\draw[fill=white](2.5,0) circle (0.1) node[below = 0.25]{\small{$6$}};
			\draw[fill=white](3,0) circle (0.1) node[below = 0.25]{\small{$0$}};
			\draw[fill=white](3.5,0) circle (0.1) node[below = 0.25]{\small{\tC}};
			\draw[fill=white](0.5,0.5) circle (0.1) node[left = 0.25]{\small{$7$}};
			\draw[fill=white](0.5,1) circle (0.1) node[left = 0.25]{\small{$8$}};
			\draw(2,1)node{$\Gamma_M$};
		\end{scope}
	\end{tikzpicture}
	\caption{Generalized Dynkin diagrams for the Narain lattice $\Gamma_N \simeq \Gamma_{1,17}$ and the Mikhailov lattice $\Gamma_M \simeq \Gamma_{9,1}$ in nine dimension.}\label{fig:gdd1}
\end{figure}
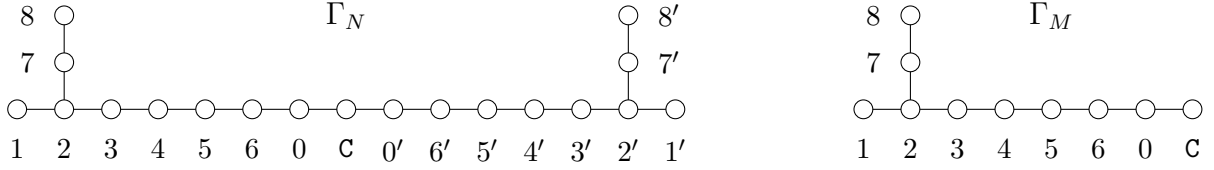

This same result can be obtained in a more concrete way by considering the GDDs for the lattices $\Gamma_N$ and $\Gamma_M$, shown in Figure \ref{fig:gdd1}. Gauge algebras in the Narain moduli space can be obtained by deleting two or more nodes of the diagram such that the result is the Dynkin diagram for an ADE root lattice. The same applies to the CHL component, but the minimum number of nodes one can delete is one. As we can see, deleting the node $0'$ in the GDD for $\Gamma_N$ gives the GDD for $\Gamma_M$ accompanied by an $\mre_8$ Dynkin diagram, from which it follows that the gauge algebras that can be obtained in each moduli space component are related as deduced above. As commented, all of the relevant groups are simply-connected. 

\subsubsection{$d = 2$}
The map from the Narain to the CHL components of the moduli space for $d = 2$ was obtained at the level of the full gauge groups in \cite{Cvetic:2021sjm} using more group-theoretical language, and proven explicitly by projecting the cocharacter lattice, which determines the topology, from $\Gamma_N$ to $\Gamma_M$. Here we briefly explain how it can be obtained in the framework of this paper.

In eight dimensions we have $\Gamma_M \simeq \Gamma_{2,2}\oplus \mrd_8$ and $\Lambda = \mrd_8$. We will consider a lattice $M$ primitively embedded into $\Gamma_N$, which is the overlattice of a root lattice $L$, containing in turn a primitively embedded $\mrd_8$ lattice. This condition restricts $L$ to be of the form
\begin{equation}
	L \simeq \mrd_{8+n}\oplus N\,,
\end{equation} 
where $n$ is some non-negative integer and $N$ is some other ADE lattice. The orthogonal complement of $\mrd_8$ in $L$ is of the form $\mrd_n\oplus N$, and has an overlattice $M'$ primitively embedded into $\Gamma_M$. 

The question is if $\mrd_n \oplus N$ is the maximal root sublattice of $M'$, according to the definition of roots in the CHL moduli space. This can indeed be verified for all points of symmetry enhancement. The subtlety here is that as lattices, $\mrd_n$ and $\mrc_n$ are equivalent. The actual contribution to the gauge algebra depends on which vectors correspond to massless states, and we find that in this case it is actually $\mathfrak{sp}_n$ and not $\mathfrak{so}_{2n}$. We therefore write $L = \mrc_n \oplus N$. We have then a simple rule for mapping gauge algebras from the Narain component to the CHL component of the moduli space. Just take any gauge algebra with a $\mrd_{8+n}$ factor and replace it with $\mrc_n$. Since it is possible to have gauge algebras with terms $\mrd_{8+n}\oplus \mrd_{8+m}$, with $m \neq n$, this map is generically one-to-many. 

To promote this map to one at the level of groups we compute the fundamental group of the gauge group associated to the embeddings $L \hookrightarrow \Gamma_N$ and $L \hookrightarrow \Gamma_M$ using the lattice methods outlined above, and then see how they are related. Each fundamental group is specified by a set of elements $\{k_i\}$ of the center of the universal cover, $Z(\tilde G)$, each one of the form $k_i = (k_i^1,...,k_i^s)$, where $k_i^j$ denotes the contribution of each of the $s$ simple factors in $\tilde G$ to $k_i$. For $\mrd_{2n}$ factors we write the corresponding contribution as a pair $(p,q) \in \mathbb{Z}_2\times \mathbb{Z}_2$. In this case we can separate each $k_i$ into the contribution of the $\mrd_{8+n}$ factor to be replaced and that of the remaining factor given by the lattice $N$. Let us write $k_i = (k_i{}^N,k_i{}^S)$ for the gauge group in the Narain component and $k'_i = (k'_i{}^N,k'_i{}^S)$ for the associated one in the CHL component. We find that for $n$ even,
\begin{equation}
	k_i{}^S = (p,q) \to k'_i{}^S = p+q \mod 2\,,
\end{equation}
while for $n$ odd
\begin{equation}
	k_i{}^S = p \to k_i'{}^S = p \mod 2 \,,
\end{equation}
while $k_i'{}^N = k_i{}^N$ in both cases. If $n = 0$, one just deletes $k_i{}^S$. 

As an example, consider the gauge group $\frac{\sping(32)}{\mathbb{Z}_2} \times \sug(2)^2$, whose fundamental group is generated by only one element $k = ((1,0),0,0)$. Using the rules above, the associated gauge group in the CHL component is $\frac{\spg(8)}{\mathbb{Z}_2} \times \sug(2)^2$ with $k = (1,0,0)$. If we had the gauge group $\sog(32)\times \sug(2)^2$ with $k = ((1,1),0,0)$, it would map to the simply-connected $\spg(8)\times \sug(2)^2$. However, $\sog(2n)$ factors are not present in the theory so that this last example does not arise. It's interesting to note that $\sog(2n)$ would map to the same gauge group as $\sping(2n)$, making the mapping generically many-to-many and not one-to-many.

Note also that the fundamental group of any two groups connected by this mapping are isomorphic. This is in accordance with the fact that the topology of the gauge groups in the dual frame of F-theory on elliptically fibered K3 is given by the torsional part of the Mordell-Weil group\cite{Aspinwall:1998xj,Mayrhofer:2014opa} which can be obtained from the Weierstrass model for the fibration (see e.g. \cite{Cvetic:2018bni}), as the mechanism of singularity freezing does not alter the Weierstrass model itself \cite{Bhardwaj:2018jgp}.

\subsubsection{$d = 3$}

Let us now review the generalization of the above results to $d = 3$ \cite{Fraiman:2021soq}. In seven dimensions we have $\Lambda = \mrd_4 \oplus \mrd_4$. Each $\mrd_4$ factor can be contained in algebras of $\mrd_{4+n}$ type, in which case the analysis for $d = 2$ goes through, including the way in which the contribution of these factors to the fundamental group transform. The difference now is that we have two such factors transforming simultaneously, e.g. $\mrd_{n+4}\oplus \mrd_{m+4} \to \mrc_n \oplus \mrc_m$. This is not the only possibility, however. 

It is also possible for $\mrd_4$ to be primitively embedded into $\mre_6,~\mre_7$ and $\mre_8$. Taking the orthogonal complement of $\mrd_4$ in each case we obtain the lattices $\mra_2(2),~3\mra_1 \simeq \mrb_3$ and $\mrd_4 \simeq \mrf_4$, respectively. Similarly to the previous case, we can look at the points of symmetry enhancement in the CHL component and determine that the contributions to the algebra are respectively $\mathfrak{su}_3$, $\mathfrak{so}_7$ and $\mathfrak{f}_4$, hence the use of these lattice isomorphisms. With respect to the gauge group's topology, we have that $Z(\sug(3)) \simeq Z(\mre_6) \simeq \mathbb{Z}_3$, $Z(\sping(7))\simeq Z(\mre_7) \simeq \mathbb{Z}_2$ and $Z(\mrf_4) \simeq Z(\mre_8) \simeq \{0\}$, and that the contributions of these factors to the $\{k_i\}$ remain invariant. This means that as for $d = 2$, the fundamental group of two gauge groups related by the mapping are isomorphic. As in the previous case, this coincides at the algebra level with results on the dual geometrical frame's mechanism of singularity freezing \cite{Atiyah:2001qf,deBoer:2001wca,Tachikawa:2015wka}, in this case M-theory on K3 with two $\mrd_4$ frozen singularities. We are not aware of how the fundamental group of the gauge group  is encoded in the M-theory compactification, but it should in any case be invariant under singularity freezing.

\subsection{Algebra projection}

In the previous constructions we have seen that the root system of the CHL gauge algebra corresponds to a subset of the orthogonal complement lattice of $\Lambda$ in the root lattice $L'$. This algebra is determined precisely by checking each case algorithmically and the result is seen to correspond to a simple general rule. Now we give a procedure whose result predicts this algebra directly, mapping the simple roots of $\mathfrak{g}_\text{Narain}$ to those of $\mathfrak{g}_\text{CHL}$. This procedure gives the correct results for $d = 1,2,3,4$· We will illustrate it case by case starting with $d = 2$, which exhibits the non-trivial features that generalize to larger $d$.

\subsubsection{$d = 2$}

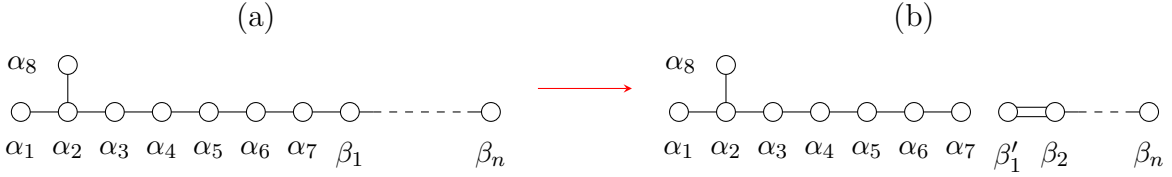
\begin{figure}
	\centering
	\begin{tikzpicture}[scale = 1.25]
		\draw(2.5,1)node{(a)};
		\draw(0,0)--(3.75,0);
		\draw[dashed](3.75,0)--(4.75,0);
		\draw(4.75,0)--(5,0);
		\draw(0.5,0)--(0.5,0.5);
		\draw[fill=white](0,0) circle (0.1) node[below = 0.25]{\small{$\alpha_1$}};
		\draw[fill=white](0.5,0) circle (0.1) node[below = 0.25]{\small{$\alpha_2$}};
		\draw[fill=white](1,0) circle (0.1) node[below = 0.25]{\small{$\alpha_3$}};
		\draw[fill=white](1.5,0) circle (0.1) node[below = 0.25]{\small{$\alpha_4$}};
		\draw[fill=white](2,0) circle (0.1) node[below = 0.25]{\small{$\alpha_5$}};
		\draw[fill=white](2.5,0) circle (0.1) node[below = 0.25]{\small{$\alpha_6$}};
		\draw[fill=white](3,0) circle (0.1) node[below = 0.25]{\small{$\alpha_7$}};
		\draw[fill=white](3.5,0) circle (0.1) node[below = 0.25]{\small{$\beta_1$}};
		\draw[fill=white](5,0) circle (0.1) node[below = 0.25]{\small{$\beta_n$}};
		\draw[fill=white](0.5,0.5) circle (0.1) node[left = 0.25]{\small{$\alpha_8$}};
		
		\draw[red,->,>=stealth](5.5,0.25)--(6.5,0.25);
		
		\begin{scope}[shift={(7,0)}]
			\draw(2.5,1)node{(b)};
			\draw(0,0)--(3,0);
			\draw[dashed](4.25,0)--(4.75,0);
			\draw(4,0)--(4.25,0);
			\draw(4.75,0)--(5,0);
			\draw(3.5,0.05)--(4,0.05);
			\draw(3.5,-0.05)--(4,-0.05);
			\draw(0.5,0)--(0.5,0.5);
			\draw[fill=white](0,0) circle (0.1) node[below = 0.25]{\small{$\alpha_1$}};
			\draw[fill=white](0.5,0) circle (0.1) node[below = 0.25]{\small{$\alpha_2$}};
			\draw[fill=white](1,0) circle (0.1) node[below = 0.25]{\small{$\alpha_3$}};
			\draw[fill=white](1.5,0) circle (0.1) node[below = 0.25]{\small{$\alpha_4$}};
			\draw[fill=white](2,0) circle (0.1) node[below = 0.25]{\small{$\alpha_5$}};
			\draw[fill=white](2.5,0) circle (0.1) node[below = 0.25]{\small{$\alpha_6$}};
			\draw[fill=white](3,0) circle (0.1) node[below = 0.25]{\small{$\alpha_7$}};
			\draw[fill=white](3.5,0) circle (0.1) node[below = 0.25]{\small{$\beta_1'$}};
			\draw[fill=white](4,0) circle (0.1) node[below = 0.25]{\small{$\beta_2$}};
			\draw[fill=white](5,0) circle (0.1) node[below = 0.25]{\small{$\beta_n$}};
			\draw[fill=white](0.5,0.5) circle (0.1) node[left = 0.25]{\small{$\alpha_8$}};
		\end{scope}
	\end{tikzpicture}
	\caption{(a) Primitive embedding of $\mrd_8$ in $\Gamma_{2,18}$ with simple roots $\alpha_i$ extended to $\mrd_{8+n}$ by $\beta_j$ (see eqs. \eqref{d8roots} and \eqref{betaroots}). (b) Projection of the roots $\beta_j$ to the orthogonal complement of $\mrd_8$ gives a $\mrc_n$ lattice and associated $\mathfrak{sp}_n$ algebra in the CHL component.} \label{fig:proj1}
\end{figure}

We start by considering a primitive embedding of $\Lambda = \mrd_8$ into $\Gamma_N \simeq \Gamma_{2,2}\oplus \Gamma_{16}$, where $\Gamma_{16}$ is the weight lattice of $\frac{\sping(32)}{\mathbb{Z}_2}$. This description makes calculations easier because $\mrd_8$ embeds primitively into $\Gamma_{16}$ but not into $\mre_8\oplus \mre_8$. A particularly simple embedding is
\begin{equation}\label{d8roots}
	\begin{split}
		\alpha_i &= \ket{0,0,0,0;0^{i-1},1,-1,0^{15-i}}\,, ~i = 1,...,7\,,\\
		\alpha_8 &= \ket{0,0,0,0;-1,-1,0^{14}}\,,
	\end{split}
\end{equation}  
where the first four entries correspond to the $\Gamma_{2,2}$ part and the other 16 to $\Gamma_{16}$. Suppose the associated gauge algebra is enhanced to $\mrd_{8+n}$ by adding $n$ simple $\beta_1,...,\beta_n$ roots forming an $\mra_n$ chain, with $\beta_1\cdot \alpha_7 = -1$. For example, take
\begin{equation}\label{betaroots}
	\beta_i = \ket{0,0,0,0;0^{i+6},1,-1,0^{8-i}}\,, ~~~ i = 1,...,n \leq 8.
\end{equation}
We will take the projection of the roots $\beta_i$ along the space orthogonal to $\mrd_8$. The roots $\beta_2,...,\beta_8$ are obviously invariant under this projection, but $\beta_1$ gets projected as
\begin{equation}
	\beta_1 \to  \ket{0,0,0,0;0^8,-1,0^7}.
\end{equation}
However, this projection is not in $\Gamma_N$, and so we multiply it by 2 to get a simple root $\beta_1' = \ket{0,0,0,0;0^8,-2,0^7}$. We see then that the simple roots of the $\mra_n$ chain get projected into the simple roots of $\mrc_n$. This construction is represented in Figure \ref{fig:proj1}, and applies to any other primitive embedding of $\mrd_8$ since it is unique up to automorphisms of $\Gamma_N$.

\subsubsection{$d = 3$}

For $d = 3$ we have $\Lambda = \mrd_4 \oplus \mrd_4$, which has an easily describable primitive embedding into $\mre_8 \oplus \mre_8$, so we use the basis $\Gamma_N \simeq \Gamma_{3,3} \oplus \mre_8 \oplus \mre_8$. This embedding reads
\begin{equation}
	\begin{split}
		\alpha_1 &= \ket{0,0,0,0;1,-1,0^6;0^8}\,, ~~~~~~~ \alpha_2 = \ket{0,0,0,0;0,1,-1,0^5;0^8}\,,\\
		\alpha_3 &= \ket{0,0,0,0;0^2,1,-1,0^4;0^8}\,, ~~~ \alpha_4 = \ket{0,0,0,0;-1,-1,0^6;0^8}\,,\\
		\alpha_1' &= \ket{0,0,0,0;0^8;1,-1,0^6}\,, ~~~~~~~ \alpha_2' = \ket{0,0,0,0;0^8;0,1,-1,0^5}\,,\\
		\alpha_3' &= \ket{0,0,0,0;0^8;0^2,1,-1,0^4}\,, ~~~ \alpha_4' = \ket{0,0,0,0;0^8;-1,-1,0^6}\,.
	\end{split}
\end{equation}
As in the previous case, we can extend each $\mrd_4$ to $\mrd_{4+n}$ with an $\mra_n$ chain, which gets projected to the orthogonal complement of $\Lambda$ as a $\mrc_n$. However, $\mrd_4$ can also be extended to $\mre_8$ passing through $\mrd_5$, $\mre_6$ and $\mre_7$. This $\mrd_5$ coincides with that of the generic extension $\mrd_{4+n}$ with $n = 1$, and so it gives rise to an $\mra_1(2)$ lattice with simple root, say, 
\begin{equation}
	\beta_1' = \ket{0,0,0,0;0,0,0,0,-2,0^3;0^8}\,,
\end{equation}
which arises from projecting $\ket{0,0,0,0;0,0,0,1,-1,0^3;0^8}$. Extending $\mrd_5$ to $\mre_6$ can be done by adding the root $\ket{0,0,0,0;\frac12^8,0^8}$. Its projection multiplied by 2 is
\begin{equation}
	\beta_2' = \ket{0,0,0,0;0^4,1^4;0^8}\,,
\end{equation}
and so we see that $\beta_1'$ and $\beta_2'$ give rise to an $\mra_2(2)$ lattice, as expected. We can further add the roots $\ket{0,0,0,0;0^4,1,-1,0,0}$ and $\ket{0,0,0,0;0^5,1,-1,0}$, extending $\mre_6$ to $\mre_7$ and then $\mre_8$. Since these roots are orthogonal to $\Lambda$, they are invariant under the projection and we see that they extend $\mra_2(2)$ to $\mrb_3$ and then $\mrf_4$ as predicted. 

\subsubsection{$d = 4$}
\label{sssec:d4proj}
Here we have $\Lambda = \mrd_8^*(2)$. This lattice has a root sublattice $L_\Lambda = 8\mra_1$ and can be in fact interpreted as the weight lattice of $\frac{\sug(2)^8}{\mathbb{Z}_2}$ with $\mathbb{Z}_2$ diagonal, i.e. $k = (1,...,1)$. A suitable primitive embedding of this lattice into $\Gamma_N \simeq \Gamma_{4,4}\oplus \mre_8 \oplus \mre_8$ has simple roots
\begin{equation}
	\begin{split}
		\alpha_1 &= \ket{0,0,0,0;1,-1,0^6;0^8}\,, ~~~~~~~ \alpha_2 = \ket{0,0,0,0;0,0,1,-1,0^4;0^8}\,,\\
		\alpha_3 &= \ket{0,0,0,0;0^4,1,-1,0^2;0^8}\,, ~~~ \alpha_4 = \ket{0,0,0,0;0^6,1,-1;0^8}\,,\\
		\alpha_5 &= \ket{0,0,0,0;0^8;1,-1,0^6}\,, ~~~~~~~ \alpha_6 = \ket{0,0,0,0;0^8;0,0,1,-1,0^4}\,,\\
		\alpha_7 &= \ket{0,0,0,0;0^8;0^4,1,-1,0^2}\,, ~~~ \alpha_8 = \ket{0,0,0,0;0^8;0^6,1,-1}\,.
	\end{split}
\end{equation}
The weight vector extending this root lattice to $\Lambda$ is just 
\begin{equation}\label{weightchl}
	w = \frac{1}{2}\sum_{i=1}^8 \alpha_i = \ket{0,0,0,0;1,0,1,0,1,0,1,0;1,0,1,0,1,0,1,0}\,.
\end{equation}
Requiring orthogonality with the roots is enough to get orthogonality with $\Lambda$, so we will not worry about $w$. However we note that there exists also a primitive embedding of $8\mra_1$ into $\Gamma_N$, which should not be confused with $\Lambda$. 

The first thing to note is that the lattice $\Lambda_L = 8A_1$ can be naively extended in many different ways but not all of them are allowed extensions of $\Lambda$ itself. For example, no $\mra_1$ factor can be individually extended to $\mra_2$ with a root orthogonal to the other $\mra_1$ factors. Any attempt to do this is easily seen to fail. The next logical step is to attach a root to two $\mra_1$ factors at the same time, e.g. with $\ket{0,0,0,0;0,1,-1,0^5;0^8}$, in this case giving an $\mra_3$. This vector gets projected to 
\begin{equation}
	\beta = \ket{0,0,0,0;1,1,-1,-1,0^4;0^8}\,,
\end{equation}
and so we have that $\mra_3$ freezes to\footnote{It is more precise to say that $\mra_3\oplus6\mra_1$ freezes to $\mra_1(2)$, but we are now focusing on the behaviour under projection of sublattices corresponding to simple algebras and not the whole lattice containing $\Lambda$.} $\mra_1(2)$. This is equivalent to $\mrd_3 \to \mrc_1$, and forms part of the more general rule $\mrd_{2+n}\to \mrc_n$, or $\mathfrak{so}_{2n+4} \to \mathfrak{sp}_n$, in analogy with those we have for $d = 2,3$. This is depicted as
\begin{eqnarray}\label{diagD2n}
	\begin{aligned}
		\begin{tikzpicture}[scale = 1.25]
			\draw(0,0)--(0.25,0);
			\draw[dashed](0.25,0)--(1.25,0);
			\draw(1.25,0)--(1.5,0);
			\draw(0,0)--(-0.5,0.5);
			\draw(0,0)--(-0.5,-0.5);
			\draw[fill=yellow](-0.5,0.5) circle (0.1) node[left = 0.25]{\small{$\alpha_1$}};
			\draw[fill=yellow](-0.5,-0.5) circle (0.1) node[left = 0.25]{\small{$\alpha_2$}};
			
			\draw[fill=white](0,0) circle (0.1) node[below = 0.25]{\small{$\beta_1$}};
			\draw[fill=white](1.5,0) circle (0.1) node[below = 0.25]{\small{$\beta_n$}};

			\draw[red,->,>=stealth](2.25,0)--(3.25,0);
			
			\begin{scope}[shift={(0.5,0)}]
				\draw[dashed](4.25,0)--(4.75,0);
				\draw(4,0)--(4.25,0);
				\draw(4.75,0)--(5,0);
				\draw(3.5,0.05)--(4,0.05);
				\draw(3.5,-0.05)--(4,-0.05);
				\draw[fill=white](3.5,0) circle (0.1) node[below = 0.25]{\small{$\beta_1'$}};
				\draw[fill=white](4,0) circle (0.1) node[below = 0.25]{\small{$\beta_2$}};
				\draw[fill=white](5,0) circle (0.1) node[below = 0.25]{\small{$\beta_n$}};
			\end{scope}
		\end{tikzpicture}
	\end{aligned}
\end{eqnarray}

The next possibility is to attach $n-1$ roots to $n$ $\mra_1$ factors in pairs such that one gets an $\mra_{2n-1}$ chain. The case $\mra_3 \to \mra_1(2)$ above can be generalized e.g. to $\mra_5 \to \mra_2(2)$ with roots
\begin{equation}
	\beta_1 = \ket{0,0,0,0;1,1,-1,-1,0^4;0^8}\,, ~~~~~ \beta_2 = \ket{0,0,0,0;0^2,1,1,-1,-1,0^2;0^8}\,,
\end{equation}
and more generally we find the rule $\mra_{2n-1} \to \mra_{n-1}(2)$, or $\mathfrak{su}_{2n} \to \mathfrak{su}_{n}$, depicted as
\begin{eqnarray}
	\begin{aligned}
		\begin{tikzpicture}[scale = 1.25]
			\draw(0,0)--(1.25,0);
			\draw(2.25,0)--(3.5,0);
			\draw[dashed](1.25,0)--(2.25,0);
			\draw[fill=yellow](0,0) circle (0.1) node[below = 0.25]{\small{$\alpha_1$}};
			\draw[fill=white](0.5,0) circle (0.1) node[above = 0.25]{\small{$\beta_1$}};
			\draw[fill=yellow](1,0) circle (0.1) node[below = 0.25]{\small{$\alpha_2$}};
			\draw[fill=yellow](2.5,0) circle (0.1) node[below = 0.25]{\small{$\alpha_{n-1}$}};
			\draw[fill=white](3,0) circle (0.1) node[above = 0.25]{\small{$\beta_{n-1}$}};
			\draw[fill=yellow](3.5,0) circle (0.1) node[below = 0.25]{\small{$\alpha_n$}};
			
			\draw[red,->,>=stealth](4.25,0)--(5.25,0);
			
			\begin{scope}[shift={(2.5,0)}]
				\draw[dashed](4.25,0)--(4.75,0);
				\draw(4,0)--(4.25,0);
				\draw(4.75,0)--(5,0);
				\draw(3.5,0)--(4,0);
				\draw[fill=white](3.5,0) circle (0.1) node[below = 0.25]{\small{$\beta_1'$}};
				\draw[fill=white](4,0) circle (0.1) node[below = 0.25]{\small{$\beta_2'$}};
				\draw[fill=white](5,0) circle (0.1) node[below = 0.25]{\small{$\beta_{n-1}'$}};
			\end{scope}
		\end{tikzpicture}
	\end{aligned}
\end{eqnarray}
From this rule we can actually get another by simply attaching a root $\beta_n$ to $\beta_{n-1}$, namely $\mrd_{2n}\to \mrb_{n}$, or $\mathfrak{so}_{4n} \to \mathfrak{so}_{2n+1}$,
\begin{eqnarray}\label{diagD2n'}
	\begin{aligned}
		\begin{tikzpicture}[scale = 1.25]
			\draw(0,0)--(1.25,0);
			\draw(2.25,0)--(3.5,0);
			\draw[dashed](1.25,0)--(2.25,0);
			\draw(3,0)--(3,-0.5);
			\draw[fill=yellow](0,0) circle (0.1) node[below = 0.25]{\small{$\alpha_1$}};
			\draw[fill=white](0.5,0) circle (0.1) node[above = 0.25]{\small{$\beta_1$}};
			\draw[fill=yellow](1,0) circle (0.1) node[below = 0.25]{\small{$\alpha_2$}};
			\draw[fill=yellow](2.5,0) circle (0.1) node[below = 0.25]{\small{$\alpha_{n-1}$}};
			\draw[fill=white](3,0) circle (0.1) node[above = 0.25]{\small{$\beta_{n-1}$}};
			\draw[fill=yellow](3.5,0) circle (0.1) node[below = 0.25]{\small{$\alpha_n$}};
			\draw[fill=white](3,-0.5) circle (0.1) node[below = 0.25]{\small{$\beta_n$}};
			
			\draw[red,->,>=stealth](4.25,0)--(5.25,0);
			
			\begin{scope}[shift={(2.5,0)}]
				\draw(5,0.05)--(5.5,0.05);
				\draw(5,-0.05)--(5.5,-0.05);
				\draw(5.20,0.15)--(5.30,0);
				\draw(5.20,-0.15)--(5.30,0);
				\draw[dashed](4.25,0)--(4.75,0);
				\draw(4,0)--(4.25,0);
				\draw(4.75,0)--(5,0);
				\draw(3.5,0)--(4,0);
				\draw[fill=white](3.5,0) circle (0.1) node[below = 0.25]{\small{$\beta_1'$}};
				\draw[fill=white](4,0) circle (0.1) node[below = 0.25]{\small{$\beta_2'$}};
				\draw[fill=white](5,0) circle (0.1) node[below = 0.25]{\small{$\beta_{n-1}'$}};
				\draw[fill=white](5.5,0) circle (0.1) node[below = 0.25]{\small{$\beta_{n}$}};
			\end{scope}
		\end{tikzpicture}
	\end{aligned}
\end{eqnarray}
Finally, we can take the particular case $n = 4$ and attach a root to $\beta_3$ to get the rule $\mre_7 \to \mrf_4$, or $\mathfrak{e}_7 \to \mathfrak{f}_4$,
\begin{eqnarray}
	\begin{aligned}
		\begin{tikzpicture}[scale = 1.25]
			\draw(0,0)--(1.25,0);
			\draw(1.25,0)--(2,0);
			\draw(1.5,0)--(1.5,-1);

			\draw[fill=yellow](0,0) circle (0.1) node[above = 0.25]{\small{$\alpha_1$}};
			\draw[fill=white](0.5,0) circle (0.1) node[above = 0.25]{\small{$\beta_1$}};
			\draw[fill=yellow](1,0) circle (0.1) node[above = 0.25]{\small{$\alpha_2$}};
			\draw[fill=white](1.5,0) circle (0.1) node[above = 0.25]{\small{$\beta_{2}$}};
			\draw[fill=yellow](2,0) circle (0.1) node[above = 0.25]{\small{$\alpha_4$}};
			\draw[fill=white](1.5,-0.5) circle (0.1) node[right = 0.25]{\small{$\beta_3$}};
			\draw[fill=white](1.5,-1) circle (0.1) node[right = 0.25]{\small{$\beta_4$}};
			
			\draw[red,->,>=stealth](3.25,0)--(4.25,0);
			
			\begin{scope}[shift={(1.5,0)}]
				\draw(4,0.05)--(4.5,0.05);
				\draw(4,-0.05)--(4.5,-0.05);
				\draw(4.20,0.15)--(4.30,0);
				\draw(4.20,-0.15)--(4.30,0);
				\draw(3.5,0)--(4,0);
				\draw(4.5,0)--(5,0);
				\draw[fill=white](3.5,0) circle (0.1) node[below = 0.25]{\small{$\beta_1'$}};
				\draw[fill=white](4,0) circle (0.1) node[below = 0.25]{\small{$\beta_2'$}};
				\draw[fill=white](4.5,0) circle (0.1) node[below = 0.25]{\small{$\beta_3$}};
				\draw[fill=white](5,0) circle (0.1) node[below = 0.25]{\small{$\beta_4$}};
			\end{scope}
		\end{tikzpicture}
	\end{aligned}
\end{eqnarray}

In summary we have found the following freezing rules at the level of the algebras:
\begin{eqnarray}\label{chlrules}
	\begin{aligned}
		\mathfrak{so}_{2n+4} &\to& \mathfrak{sp}_n\,\\
		\mathfrak{su}_{2n} &\to& \mathfrak{su}_{n}\,\\	
		\mathfrak{so}_{4n} &\to& \mathfrak{so}_{2n+1}\,\\
		\mathfrak{e}_7 &\to& \mathfrak{f}_4\,
	\end{aligned}
\end{eqnarray}
where both the LHS and RHS algebras are at level 1 (the algebras unaffected by the freezing become level 2). These rules cannot be applied arbitrarily, however. In order for the LHS algebras to be reduced to those in the RHS, their roots must be connected with those of $\Lambda$ as specified in each case above. Any root of $\Lambda$ left by itself is simply projected out, $\mathfrak{su}_2 \to \emptyset$.

\subsection{Applying the map in \texorpdfstring{$d = 4$}{d=4}}
\label{ssec:chl4d}
Having seen the possible ways in which subalgebras of a gauge algebra in the Narain component in six dimensions can be transformed when mapping to the CHL component, we now treat the problem of when these rules are applicable for a given gauge group $G$. In the cases $d=1,2,3$ this problem is trivial because the root lattices associated to $\Lambda$ are uniquely embedded, so one always knows for any gauge group if its weight lattice contains $\Lambda$ by a simple reading of the algebra. For $d = 4$, however, the relevant root lattice is $8 \mra_1$, which may or may not be associated to $\Lambda$. It is necessary therefore to check explicitly, for each $8\mra_1$ sublattice, if it corresponds to $\Lambda$ or not.

As a simple example let us consider the gauge group $\frac{\sping(32)}{\mathbb{Z}_2}$, ignoring the extra four $\uo$ factors for now. It turns out that the weight lattice of this group contains $\Lambda$ as a sublattice, whose $8\mra_1$ sublattice correspond to the yellow nodes in the diagram
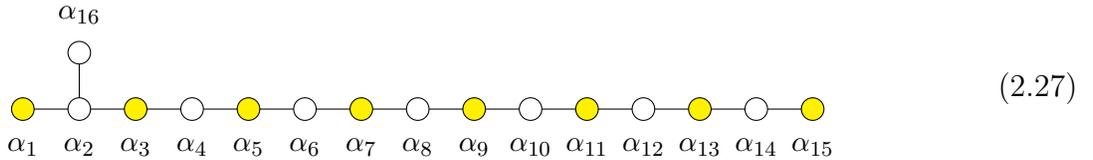
\begin{eqnarray}
	\begin{aligned}
		\begin{tikzpicture}[scale = 1.5]
			\draw(0,0)--(7,0);
			\draw(0.5,0)--(0.5,0.5);
			\draw[fill=yellow](0,0) circle (0.1) node[below = 0.25]{\small{$\alpha_1$}};
			\draw[fill=white](0.5,0) circle (0.1) node[below = 0.25]{\small{$\alpha_2$}};
			\draw[fill=yellow](1,0) circle (0.1) node[below = 0.25]{\small{$\alpha_3$}};
			\draw[fill=white](1.5,0) circle (0.1) node[below = 0.25]{\small{$\alpha_4$}};
			\draw[fill=yellow](2,0) circle (0.1) node[below = 0.25]{\small{$\alpha_5$}};
			\draw[fill=white](2.5,0) circle (0.1) node[below = 0.25]{\small{$\alpha_6$}};
			\draw[fill=yellow](3,0) circle (0.1) node[below = 0.25]{\small{$\alpha_7$}};
			\draw[fill=white](3.5,0) circle (0.1) node[below = 0.25]{\small{$\alpha_8$}};
			\draw[fill=yellow](4,0) circle (0.1) node[below = 0.25]{\small{$\alpha_9$}};
			\draw[fill=white](4.5,0) circle (0.1) node[below = 0.25]{\small{$\alpha_{10}$}};
			\draw[fill=yellow](5,0) circle (0.1) node[below = 0.25]{\small{$\alpha_{11}$}};
			\draw[fill=white](5.5,0) circle (0.1) node[below = 0.25]{\small{$\alpha_{12}$}};
			\draw[fill=yellow](6,0) circle (0.1) node[below = 0.25]{\small{$\alpha_{13}$}};
			\draw[fill=white](6.5,0) circle (0.1) node[below = 0.25]{\small{$\alpha_{14}$}};
			\draw[fill=yellow](7,0) circle (0.1) node[below = 0.25]{\small{$\alpha_{15}$}};
			\draw[fill=white](0.5,0.5) circle (0.1) node[above = 0.25]{\small{$\alpha_{16}$}};
		\end{tikzpicture}
	\end{aligned}
\end{eqnarray}
This can be shown explicitly by deleting the white nodes and checking that the weight vector 
of the $\frac{\sug(2)^8}{\mathbb{Z}_2}$ weight lattice is in the Narain lattice (cf. eq. \eqref{weightchl}). At the level of the algebras, then, we have that $\mathfrak{so}_{32}$ goes to $\mathfrak{so}_{17}$. This is to be contrasted with the gauge group $\sping(32)$, which is simply-connected and therefore does not contain $\Lambda$ in its weight lattice (which is a root lattice in this case). From this we learn that the topology of the group dictates what are the allowed freezings. Furthermore, we can explicitly compute the fundamental group of the gauge groups using the methods of \cite{Fraiman:2021soq}, which extend to $d = 4$, and find that $\frac{\sping(32)}{\mathbb{Z}_2}$ gets mapped to $\sping(17)$. In other words, the element $k' = (1,0)$ which generates $\pi_1(\frac{\sping(32)}{\mathbb{Z}_2})$ gets mapped to $k = 0$ in $\pi_1(\sping(17))$. 
%
%

In general, the gauge group to be mapped has more than one nontrivial element in its fundamental group, which makes things more complicated. Consider for example
\begin{equation}
	G = \frac{\sug(2)\times \sug(4) \times \sug(4) \times \sping(12) \times \mre_7}{\mathbb{Z}_2 \times \mathbb{Z}_2}\,,
\end{equation}
where the $\mathbb{Z}_2\times \mathbb{Z}_2$ consists of 
\begin{equation}
	k_1 = (0,2,0,(1,0),1)\,, ~~~~~~~ k_2 = (1,0,2,(1,1),1) \,, ~~~~~~~ k_3 = (1,2,2,(0,1),0)\,.
\end{equation}
Any pair of these elements, which generate $\pi_1(G)$, corresponds to two vectors which extend the root lattice $L$ of $G$ to its weight lattice $M$. They are inequivalent under translations in $L$. One can then delete nodes in the Dynkin diagram of $L$ such that the reduced root lattice still has a nontrivial weight overlattice which might correspond to $\Lambda$, at which point any other reduction will not contain weight vectors. In this special case, all such reductions lead to inequivalent embeddings of $\Lambda$ in $M$, represented by the yellow nodes in the diagrams
\begin{eqnarray}
	\begin{aligned}
		\begin{tikzpicture}[scale = 1.5]
			\draw(0.5,0)--(1.5,0);
			\draw(2,0)--(3,0);
			\draw(3.5,0)--(5.5,0);
			\draw(5,0)--(5,0.5);
			\draw(6,0)--(8,0.0);
			\draw(7.5,0)--(7.5,1);
			\draw[fill=white](0,0) circle (0.1) node[below = 0.25]{\small{$\alpha_1$}};
			\draw[fill=yellow](0.5,0) circle (0.1) node[below = 0.25]{\small{$\alpha_2$}};
			\draw[fill=white](1,0) circle (0.1) node[below = 0.25]{\small{$\alpha_3$}};
			\draw[fill=yellow](1.5,0) circle (0.1) node[below = 0.25]{\small{$\alpha_4$}};
			\draw[fill=white](2,0) circle (0.1) node[below = 0.25]{\small{$\alpha_5$}};
			\draw[fill=white](2.5,0) circle (0.1) node[below = 0.25]{\small{$\alpha_6$}};
			\draw[fill=white](3,0) circle (0.1) node[below = 0.25]{\small{$\alpha_7$}};
			\draw[fill=yellow](3.5,0) circle (0.1) node[below = 0.25]{\small{$\alpha_8$}};
			\draw[fill=white](4,0) circle (0.1) node[below = 0.25]{\small{$\alpha_9$}};
			\draw[fill=yellow](4.5,0) circle (0.1) node[below = 0.25]{\small{$\alpha_{10}$}};
			\draw[fill=white](5,0) circle (0.1) node[below = 0.25]{\small{$\alpha_{11}$}};
			\draw[fill=yellow](5.5,0) circle (0.1) node[below = 0.25]{\small{$\alpha_{12}$}};
			\draw[fill=white](5,0.5) circle (0.1) node[right = 0.25]{\small{$\alpha_{13}$}};
			\draw[fill=yellow](6,0) circle (0.1) node[below = 0.25]{\small{$\alpha_{14}$}};
			\draw[fill=white](6.5,0) circle (0.1) node[below = 0.25]{\small{$\alpha_{15}$}};
			\draw[fill=yellow](7,0) circle (0.1) node[below = 0.25]{\small{$\alpha_{16}$}};
			\draw[fill=white](7.5,0) circle (0.1) node[below = 0.25]{\small{$\alpha_{17}$}};
			\draw[fill=yellow](8,0) circle (0.1) node[below = 0.25]{\small{$\alpha_{18}$}};
			\draw[fill=white](7.5,0.5) circle (0.1) node[right = 0.25]{\small{$\alpha_{19}$}};
			\draw[fill=white](7.5,1) circle (0.1) node[right = 0.25]{\small{$\alpha_{20}$}};
		\end{tikzpicture}
	\end{aligned}\\
	\begin{aligned}
		\begin{tikzpicture}[scale = 1.5]
			\draw(0.5,0)--(1.5,0);
			\draw(2,0)--(3,0);
			\draw(3.5,0)--(5.5,0);
			\draw(5,0)--(5,0.5);
			\draw(6,0)--(8,0.0);
			\draw(7.5,0)--(7.5,1);
			\draw[fill=yellow](0,0) circle (0.1) node[below = 0.25]{\small{$\alpha_1$}};
			\draw[fill=white](0.5,0) circle (0.1) node[below = 0.25]{\small{$\alpha_2$}};
			\draw[fill=white](1,0) circle (0.1) node[below = 0.25]{\small{$\alpha_3$}};
			\draw[fill=white](1.5,0) circle (0.1) node[below = 0.25]{\small{$\alpha_4$}};
			\draw[fill=yellow](2,0) circle (0.1) node[below = 0.25]{\small{$\alpha_5$}};
			\draw[fill=white](2.5,0) circle (0.1) node[below = 0.25]{\small{$\alpha_6$}};
			\draw[fill=yellow](3,0) circle (0.1) node[below = 0.25]{\small{$\alpha_7$}};
			\draw[fill=white](3.5,0) circle (0.1) node[below = 0.25]{\small{$\alpha_8$}};
			\draw[fill=white](4,0) circle (0.1) node[below = 0.25]{\small{$\alpha_9$}};
			\draw[fill=white](4.5,0) circle (0.1) node[below = 0.25]{\small{$\alpha_{10}$}};
			\draw[fill=white](5,0) circle (0.1) node[below = 0.25]{\small{$\alpha_{11}$}};
			\draw[fill=yellow](5.5,0) circle (0.1) node[below = 0.25]{\small{$\alpha_{12}$}};
			\draw[fill=yellow](5,0.5) circle (0.1) node[right = 0.25]{\small{$\alpha_{13}$}};
			\draw[fill=yellow](6,0) circle (0.1) node[below = 0.25]{\small{$\alpha_{14}$}};
			\draw[fill=white](6.5,0) circle (0.1) node[below = 0.25]{\small{$\alpha_{15}$}};
			\draw[fill=yellow](7,0) circle (0.1) node[below = 0.25]{\small{$\alpha_{16}$}};
			\draw[fill=white](7.5,0) circle (0.1) node[below = 0.25]{\small{$\alpha_{17}$}};
			\draw[fill=yellow](8,0) circle (0.1) node[below = 0.25]{\small{$\alpha_{18}$}};
			\draw[fill=white](7.5,0.5) circle (0.1) node[right = 0.25]{\small{$\alpha_{19}$}};
			\draw[fill=white](7.5,1) circle (0.1) node[right = 0.25]{\small{$\alpha_{20}$}};
		\end{tikzpicture}
	\end{aligned}\\
	\begin{aligned}
		\begin{tikzpicture}[scale = 1.5]
			\draw(0.5,0)--(1.5,0);
			\draw(2,0)--(3,0);
			\draw(3.5,0)--(5.5,0);
			\draw(5,0)--(5,0.5);
			\draw(6,0)--(8,0.0);
			\draw(7.5,0)--(7.5,1);
			\draw[fill=yellow](0,0) circle (0.1) node[below = 0.25]{\small{$\alpha_1$}};
			\draw[fill=yellow](0.5,0) circle (0.1) node[below = 0.25]{\small{$\alpha_2$}};
			\draw[fill=white](1,0) circle (0.1) node[below = 0.25]{\small{$\alpha_3$}};
			\draw[fill=yellow](1.5,0) circle (0.1) node[below = 0.25]{\small{$\alpha_4$}};
			\draw[fill=yellow](2,0) circle (0.1) node[below = 0.25]{\small{$\alpha_5$}};
			\draw[fill=white](2.5,0) circle (0.1) node[below = 0.25]{\small{$\alpha_6$}};
			\draw[fill=yellow](3,0) circle (0.1) node[below = 0.25]{\small{$\alpha_7$}};
			\draw[fill=yellow](3.5,0) circle (0.1) node[below = 0.25]{\small{$\alpha_8$}};
			\draw[fill=white](4,0) circle (0.1) node[below = 0.25]{\small{$\alpha_9$}};
			\draw[fill=yellow](4.5,0) circle (0.1) node[below = 0.25]{\small{$\alpha_{10}$}};
			\draw[fill=white](5,0) circle (0.1) node[below = 0.25]{\small{$\alpha_{11}$}};
			\draw[fill=white](5.5,0) circle (0.1) node[below = 0.25]{\small{$\alpha_{12}$}};
			\draw[fill=yellow](5,0.5) circle (0.1) node[right = 0.25]{\small{$\alpha_{13}$}};
			\draw[fill=white](6,0) circle (0.1) node[below = 0.25]{\small{$\alpha_{14}$}};
			\draw[fill=white](6.5,0) circle (0.1) node[below = 0.25]{\small{$\alpha_{15}$}};
			\draw[fill=white](7,0) circle (0.1) node[below = 0.25]{\small{$\alpha_{16}$}};
			\draw[fill=white](7.5,0) circle (0.1) node[below = 0.25]{\small{$\alpha_{17}$}};
			\draw[fill=white](8,0) circle (0.1) node[below = 0.25]{\small{$\alpha_{18}$}};
			\draw[fill=white](7.5,0.5) circle (0.1) node[right = 0.25]{\small{$\alpha_{19}$}};
			\draw[fill=white](7.5,1) circle (0.1) node[right = 0.25]{\small{$\alpha_{20}$}};
		\end{tikzpicture}
	\end{aligned}
\end{eqnarray}

To each of these embeddings corresponds a different way of mapping $G$ to the CHL component. Using the rules in \eqref{chlrules} and computing the fundamental group in each case we get, respectively,
\begin{eqnarray}
	G &\to& 
	\frac{\sug(2) \times \sug(2) \times \sug(4) \times \sping(7) \times \mrf_4}{\mathbb{Z}_2}\,, ~~~ k = (1,0,2,1,0)\,,\\
	G&\to& \frac{\sug(4)\times \sug(2) \times \spg(4) \times \mrf_4}{\mathbb{Z}_2}\,,~~~~~~~~~~~~~~~~~k = (2,0,1,0)\,,\\
	G&\to& \frac{\sug(2) \times \sug(2) \times \sping(7) \times \mre_7}{\mathbb{Z}_2}\,, ~~~~~~~~~~~~~~ k = (0,0,1,1)\,.
\end{eqnarray}
The first thing to note is that in the resulting gauge group the fundamental group always reduces by a factor of $\mathbb{Z}_2$ (as already happened in the $\frac{\sping(32)}{\mathbb{Z}_2} \to \sping(17)$ case above). This can be understood by noting that one is taking the orthogonal complement of $\Lambda$, which contains weight vectors. These are also weight vectors in $M$, equivalent under translations in $8\mra_1$, so they can be related to one of the elements in $\pi_1(G)$. For any such weight vector $w$, we have that $2w \in 8\mra_1$ and so the associated $k \in \pi_1(G)$ generates a $\mathbb{Z}_2$. This is precisely the factor which is eliminated in mapping $G$, corresponding respectively to $k_1$, $k_2$ and $k_3$ above. 

Now we need to know how the remaining $k$'s get transformed in each case. What we find is that it suffices to mod every $k$ by the one that is eliminated, call it $k_\Lambda$,
\begin{equation}
	k \to k \mod k_\Lambda\,
\end{equation}
and then project it into the center of the resulting gauge group. In the case of a $\sping(4n)$ factor, we project the modded $k$ contribution to $1 \in \pi_1(\sping(2n+1)) = \mathbb{Z}_2$ if it is not $(0,0)$. Of course, we also have that $k_\Lambda \to 0$ so that this rule applies equally well to all the $k$'s of $\pi_1(G)$. 

We see then that the only information we require to know how to map a group $G$ to the CHL component is the embedding of the roots of $\Lambda$ into the root lattice $L$ of $G$ and its associated $k_\Lambda \in \pi_1(G)$. In fact, however, these two pieces of data are the same. One can take any $k \in \pi_1(G)$ of order 2 in  $Z(G)$ and check if it corresponds to $\Lambda$ in the following way. For each simple factor in $G$, if the corresponding entry in $k$ is nonzero, its Dynkin diagram should be labeled according to one of the diagrams of Section \ref{sssec:d4proj}. The only simple factor which contains more than one order 2 element is $\mrd_{2n}$, in which case $k_{D_{2n}} = (1,1)$ corresponds to the diagram \eqref{diagD2n} and $k_{D_{2n}} = (0,1), (1,0)$ correspond to \eqref{diagD2n'}. Coloring the nodes appropriately lead to those shown in the example above, as one can easily check. If there are in total eight yellow nodes, this labeling will correspond to an embedding of $\Lambda$ into $M$. With this we can apply the mapping rules to the algebra and to the fundamental group of $G$. 

We verified all of these statements by applying the procedures outlined above to a reasonably exhaustive list of gauge groups in the Narain component, and checking the results against a list for the CHL string. In the next section we look at other rank reduced components, where the results are similarly verified against lists of symmetry enhancements. These lists can be obtained in the same way as those of the 7d case in \cite{Fraiman:2021soq}, and can be accessed at \cite{fp2021}. We provide various examples in Appendix \ref{app:examples}.  

\section{Other rank reduced components}
\label{sec:others}

In this section we extend the freezing procedure explained above to other rank reduced components in the moduli space of heterotic strings which appear in seven dimensions and below. These correspond to the holonomy triples constructed in \cite{deBoer:2001wca} and their torus compactifications. We will focus our attention on the six dimensional case. Before this, however, let us review that of seven dimensions\cite{Fraiman:2021soq}.

\subsection{Review of the map in 7d}

In seven dimensions there are six connected components in the moduli space of supersymmetric heterotic strings, including the Narain and the CHL component. They can be obtained as asymmetric orbifolds of the $T^3$ compactifications at points in the moduli space where the Narain lattice exhibits appropriate symmetries. These orbifolds are of order $2$ to $6$, and they correspond to non-trivial holonomy triples in the target space, hence they are called $\mathbb{Z}_n$-triples with $n = 2,...,6$. The $\mathbb{Z}_2$-triple is equivalent to the CHL string treated in Section \ref{sec:CHL}.  Let us then treat the cases $n = 3,4,5,6$. 

For each $\mathbb{Z}_n$-triple, the momentum lattice can be obtained as the orthogonal complement of some other lattice $\Lambda$. This data is shown in Table \ref{tab:lattices7d}.
\begin{table}
	\begin{center}
		\begin{tabular}{|c|c|c|}\hline
			$n$ & Momentum lattice $\Gamma$ & Frozen sublattice $\Lambda$ \\ \hline
			1&$\rii_{3,3}\oplus \mre_8 \oplus \mre_8$&$\emptyset$\\  \hline
			2&$\rii_{3,3} \oplus  \mrf_4 \oplus \mrf_4$&$\mrd_4\oplus \mrd_4$\\ \hline
			3&$\rii_{3,3}\oplus \mrg_2 \oplus \mrg_2$&$\mre_6 \oplus \mre_6$\\ \hline
			4&$\rii_{3,3}\oplus \mra_1 \oplus \mra_1$&$\mre_7 \oplus \mre_7$\\ \hline
			5&$\rii_{3,3}$&$\mre_8\oplus \mre_8$\\\hline
			6&$\rii_{3,3}$&$\mre_8 \oplus \mre_8$ \\ \hline
		\end{tabular}
		\caption{Momentum lattices for the moduli spaces of heterotic $\mathbb{Z}_m$-triples and their orthogonal complements $\Lambda$ in $\Gamma_{3,19}$.}
		\label{tab:lattices7d}
	\end{center}
\end{table}
For the $\mathbb{Z}_3$-triple, we have $\Lambda = \mre_6 \oplus \mre_6$, which can only be embedded into $\mre_p \oplus \mre_q$ with $p,q = 6,7,8$. For each $\mre_p$ factor, we have the algebra mapping
\begin{equation}
	\mathfrak{e}_6 \to \emptyset\,, ~~~~~ \mathfrak{e}_7 \to \mathfrak{su}_2\,, ~~~~~ \mathfrak{e}_8 \to \mathfrak{g}_2\,,
\end{equation}
while the corresponding contribution to any element $k$ of the fundamental group is preserved. As with the $n = 2$ component, we have that the gauge groups related by the mapping have isomorphic fundamental groups. 

For the $\mathbb{Z}_4$-triple, we have $\Lambda = \mre_7 \oplus \mre_7$, which embeds only into $\mre_p \oplus \mre_q$ with $p,q = 7,8$. For each $\mre_p$ factor we have the algebra mapping
\begin{equation}
	\mathfrak{e}_7 \to \emptyset\,, ~~~~~ \mathfrak{e}_8 \to \mathfrak{su}_2\,.
\end{equation}

The $\mathbb{Z}_5$ and $\mathbb{Z}_6$-triples both have $\Lambda = \mre_8 \oplus \mre_8$ and so the only mapping allowed is $\mathfrak{e}_8 \to \emptyset$. All the possible gauge groups involved in this mapping are simply-connected so here again they have isomorphic fundamental groups, namely trivial ones. 

\subsection{Extension of the freezing map in 6d} 

Let us now consider the compactifications of the 7d $\mathbb{Z}_n$-triples to 6d with $n = 3,4,5,6$. Not surprisingly, the mappings that we find here generalize naturally those of the $n = 2$ case.

\subsubsection{6d $\mathbb{Z}_3$-triple}

For $n = 3$, the momentum lattice is
\begin{equation}
	\Gamma_{3,3}\oplus \Gamma_{1,1}(3)\oplus \mra_2 \oplus \mra_2\,,
\end{equation}
which can be shown to be the orthogonal complement of a lattice $\Lambda$ in $\Gamma_{4,20}$ isomorphic to the weight lattice of $\frac{\sug(3)^6}{\mathbb{Z}_3}$, with $\mathbb{Z}_3$ diagonal. There are two types of root lattices which can be obtained by attaching nodes to the Dynkin diagram of this $\sug(3)^6$. First, we have those of the type $\mra_{3n-1}$, obtained by adding roots between each pair of $\mra_2$'s consecutively. These map to $\mra_n$. For example, we have that $\mra_8 \to \mra_2$, 
\begin{eqnarray}
	\begin{aligned}
		\begin{tikzpicture}[scale = 1.25]
			\draw(0,0)--(3.5,0);
			\draw(6,0)--(6.5,0);
			\draw[fill=yellow](0,0) circle (0.1) node[below = 0.25]{\small{$\alpha_1$}};
			\draw[fill=yellow](0.5,0) circle (0.1) node[below = 0.25]{\small{$\alpha_2$}};
			\draw[fill=white](1,0) circle (0.1) node[below = 0.25]{\small{$\beta_1$}};
			\draw[fill=yellow](1.5,0) circle (0.1) node[below = 0.25]{\small{$\alpha_3$}};
			\draw[fill=yellow](2,0) circle (0.1) node[below = 0.25]{\small{$\alpha_4$}};
			\draw[fill=white](2.5,0) circle (0.1) node[below = 0.25]{\small{$\beta_2$}};
			\draw[fill=yellow](3,0) circle (0.1) node[below = 0.25]{\small{$\alpha_5$}};
			\draw[fill=yellow](3.5,0) circle (0.1) node[below = 0.25]{\small{$\alpha_6$}};
			\draw[red,->,>=stealth](4.25,0)--(5.25,0);
			\draw[fill=white](6,0) circle (0.1) node[below = 0.25]{\small{$\beta_1'$}};
			\draw[fill=white](6.5,0) circle (0.1) node[below = 0.25]{\small{$\beta_2'$}};
		\end{tikzpicture}
	\end{aligned}
\end{eqnarray}
The other possibility is to map $\mre_6$ to $\mrg_2$,
\begin{eqnarray}
	\begin{aligned}
		\begin{tikzpicture}[scale = 1.25]
			\draw(0,0)--(2,0);
			\draw(4.5,0)--(5,0);
			\draw(1,0)--(1,0.5);
			\draw(4.5,0.05)--(5,0.05);
			\draw(4.5,-0.05)--(5,-0.05);
			\draw(4.70,0.15)--(4.80,0);
			\draw(4.70,-0.15)--(4.80,0);
			\draw[fill=yellow](0,0) circle (0.1) node[below = 0.25]{\small{$\alpha_1$}};
			\draw[fill=yellow](0.5,0) circle (0.1) node[below = 0.25]{\small{$\alpha_2$}};
			\draw[fill=white](1,0) circle (0.1) node[below = 0.25]{\small{$\beta_1$}};
			\draw[fill=yellow](1.5,0) circle (0.1) node[below = 0.25]{\small{$\alpha_3$}};
			\draw[fill=yellow](2,0) circle (0.1) node[below = 0.25]{\small{$\alpha_4$}};
			\draw[fill=white](1,0.5) circle (0.1) node[right = 0.25]{\small{$\beta_2$}};
			\draw[red,->,>=stealth](2.75,0)--(3.75,0);
			\draw[fill=white](4.5,0) circle (0.1) node[below = 0.25]{\small{$\beta_1'$}};
			\draw[fill=white](5,0) circle (0.1) node[below = 0.25]{\small{$\beta_2$}};
		\end{tikzpicture}
	\end{aligned}
\end{eqnarray}
A gauge group $G$ in the Narain component can be mapped to this moduli space if $\pi_1(G)$ contains an order 3 element $k_\Lambda$ such that its entries label exactly 12 nodes in the associated Dynkin diagram, in a manner completely analogous to the case for the CHL string (see Section \ref{ssec:chl4d}). The procedure for mapping all the elements of $\pi_1(G)$ is the same. For example, the gauge group $\frac{\mre_6^3}{\mathbb{Z}_3}$ with $\pi_1$ generator $k = (1,1,1)$ maps to $\mrg_2^3$, and $\frac{\sug(3)^2 \times \sug(6)^2\times \sping(10)}{\mathbb{Z}_6}$ with $\pi_1$ generator $k = (1,1,1,1,2)$ maps to $\frac{\sug(2)^2 \times \sping(10)}{\mathbb{Z}_2}$ with $k = (1,1,2)$. Similarly to the CHL string, the unaltered simple factors correspond to level 3 algebras and the altered to level 1 ones, so that e.g. the latter has algebra $(\mathfrak{su}_2\oplus \mathfrak{su}_2)_1 \oplus (\mathfrak{spin}_{10})_3$. 

\subsubsection{6d $\mathbb{Z}_4$-triple}

For $n = 4$, the momentum lattice is $\Gamma_{3,3}\oplus \Gamma_{1,1}(4) \oplus \mra_1 \oplus \mra_1$,
whose associated $\Lambda$ is the weight lattice of $\frac{\sug(2)^2 \times \sug(4)^4}{\mathbb{Z}_4}$ with $\mathbb{Z}_4$ generated by $k = (1,1,1,1,1,1)$. The roots of this lattice can be extended in particular to $\mra_{4n-1}$ and $\mrd_{2n+3}$, the latter with $n = 1,2$. The algebras are respectively mapped to $\mathfrak{su}_n$ and $\mathfrak{sp}_n$. For example, we have
\begin{eqnarray}
	\begin{aligned}
		\begin{tikzpicture}[scale = 1.25]
			\draw(-1.5,0)--(3.5,0);
			\draw(6,0)--(6.5,0);
			\draw[fill=yellow](-1.5,0) circle (0.1) node[below = 0.25]{\small{$\alpha_1$}};
			\draw[fill=yellow](-1,0) circle (0.1) node[below = 0.25]{\small{$\alpha_2$}};
			\draw[fill=yellow](-0.5,0) circle (0.1) node[below = 0.25]{\small{$\alpha_3$}};
			\draw[fill=white](0,0) circle (0.1) node[below = 0.25]{\small{$\beta_1$}};
			\draw[fill=yellow](0.5,0) circle (0.1) node[below = 0.25]{\small{$\alpha_4$}};
			\draw[fill=yellow](1,0) circle (0.1) node[below = 0.25]{\small{$\alpha_5$}};
			\draw[fill=yellow](1.5,0) circle (0.1) node[below = 0.25]{\small{$\alpha_6$}};
			\draw[fill=white](2,0) circle (0.1) node[below = 0.25]{\small{$\beta_2$}};
			\draw[fill=yellow](2.5,0) circle (0.1) node[below = 0.25]{\small{$\alpha_7$}};
			\draw[fill=yellow](3,0) circle (0.1) node[below = 0.25]{\small{$\alpha_8$}};
			\draw[fill=yellow](3.5,0) circle (0.1) node[below = 0.25]{\small{$\alpha_9$}};
			\draw[red,->,>=stealth](4.25,0)--(5.25,0);
			\draw[fill=white](6,0) circle (0.1) node[below = 0.25]{\small{$\beta_1'$}};
			\draw[fill=white](6.5,0) circle (0.1) node[below = 0.25]{\small{$\beta_2'$}};
		\end{tikzpicture}
	\end{aligned}
\end{eqnarray}
\begin{eqnarray}
	\begin{aligned}
		\begin{tikzpicture}[scale = 1.25]
			\draw(0,0)--(2,0);
			\draw(0,0)--(-0.5,0.5);
			\draw(0,0)--(-0.5,-0.5);
			\draw[fill=yellow](-0.5,0.5) circle (0.1) node[left = 0.25]{\small{$\alpha_1$}};
			\draw[fill=yellow](-0.5,-0.5) circle (0.1) node[left = 0.25]{\small{$\alpha_2$}};
			
			\draw[fill=yellow](0,0) circle (0.1) node[below = 0.25]{\small{$\alpha_3$}};
			\draw[fill=white](0.5,0) circle (0.1) node[below = 0.25]{\small{$\beta_1$}};
			\draw[fill=yellow](1,0) circle (0.1) node[below = 0.25]{\small{$\alpha_4$}};
			\draw[fill=white](1.5,0) circle (0.1) node[below = 0.25]{\small{$\beta_2$}};
			\draw[fill=yellow](2,0) circle (0.1) node[below = 0.25]{\small{$\alpha_5$}};

			\draw[red,->,>=stealth](2.75,0)--(3.75,0);
			
			\begin{scope}[shift={(1,0)}]
				\draw(3.5,0.05)--(4,0.05);
				\draw(3.5,-0.05)--(4,-0.05);
                    \draw(3.70,0.15)--(3.80,0);
                    \draw(3.70,-0.15)--(3.80,0);
				\draw[fill=white](3.5,0) circle (0.1) node[below = 0.25]{\small{$\beta_1'$}};
				\draw[fill=white](4,0) circle (0.1) node[below = 0.25]{\small{$\beta_2'$}};
			\end{scope}
		\end{tikzpicture}
	\end{aligned}
\end{eqnarray}
In the latter case we see that the two $\mra_1$'s of $\Lambda$ are used up, so that one cannot extend to $\mrd_9$ and beyond. The resulting gauge groups have current algebras at level 1, except for the case of $\mra_3$ which only involves two frozen $\mra_1$'s and produces an $\mra_1$ at level 2. Unaffected factors become level 4.

The element $k_\Lambda$ associated to this mapping is of order 4. In particular this means that $2k_\Lambda$ is an order 2 element, which turns out to be associated to the freezing to the CHL component of the moduli space. This is reflected in the fact that the frozen sublattice of this moduli space component contains the one for the CHL component. Indeed, the $2\mra_1$ part of $L_\Lambda$ can be extended to $\mrd_n$ and frozen to $\mrc_{n-2}$, as for the CHL freezing rule. This will be the case if $k_\Lambda$ has an order 2 contribution to a $\mrd_n$ factor. 

For example, the group $\frac{\sug(2)^3 \times \sug(4) \times \sug(8)^2}{\mathbb{Z}_8}$ with $\pi_1$ generator $k=(1,1,1,1,1,1)$ maps to 
$\frac{\sug(2)^6}{\mathbb{Z}_2}$ with $\pi_1$ generator $k=(1,1,1,1,1,1)$ and algebra 
$(\mathfrak{su}_2\oplus \mathfrak{su}_2)_1 \oplus (\mathfrak{su}_2)_2\oplus (\mathfrak{su}_2\oplus \mathfrak{su}_2\oplus \mathfrak{su}_2)_4$, showcasing the possible ways in which current algebra levels can mix; here the mapping is associated to the order four element $2k \simeq (0,0,0,2,2,2)$. Another interesting example is given by the group $\sug(3) \times \frac{ \sug(12) \times \sping(14)}{\mathbb{Z}_4}$ with $\pi_1$ generator $k=(0,3,3)$, which maps to 
$\sug(3)^2 \times \spg(2)$, with algebra 
$(\mathfrak{sp}_2\oplus \mathfrak{su}_3)_1 \oplus (\mathfrak{su}_3)_4$; this involves the freezing rule for $\mrd_7$, producing a simply-connected gauge group.

\subsubsection{6d $\mathbb{Z}_{5,6}$-triples}
For $n = 5$ the momentum lattice is $\Gamma_{3,3}\oplus \Gamma_{1,1}(5)$, whose associated $\Lambda$ is the weight lattice of $\sug(5)^4/\mathbb{Z}_5$ with $\mathbb{Z}_5$ generated by $k = (1,1,1,1)$. The only extension allowed here is $\mra_{5n-1}$, which maps to $\mra_{n-1}$, generalizing the similar freezings in the previous cases. 

For $n = 6$, we have momentum lattice $\Gamma_{3,3}\oplus \Gamma_{1,1}(6)$, whose $\Lambda$ is the weight lattice of $\tfrac{\sug(2)^2\times \sug(3)^2 \times \sug(6)^2}{\mathbb{Z}_6}$
with $\mathbb{Z}_6$ generated by $(1,1,1,1,1,1)$. Again, the only allowed freezing here will be from $\mra_{6n-1}$ to $\mra_{n-1}$, associated to an order 6 element in $\pi_1(G)$. However, this $\Lambda$ includes the frozen sublattices of $n = 2$ and $n = 3$. Similarly to the $n=4$ case including $n=2$ freezing rules, here we also have the $n = 2$ and $n = 3$ rules which can be realized by two $\mra_1$ factors and two $\mra_2$ factors, respectively.

\subsection{Relation with \texorpdfstring{$G$}{G}-bundles over \texorpdfstring{$T^2$}{T2}}
So far we have shown that depending on the topology of a gauge group $G$ in the Narain component of the 6d moduli space one can map it to another gauge group $G'$ in a different component using a simple set of rules. Associated to every freezing there is an element of the fundamental group $k_\Lambda \in \pi_1(G)$, and depending on the order of its entries with respect to the center of each simple factor, the associated algebra will transform in a specific way. This rules are summarized in Table \ref{tab:rules}, where we've also indicated the contribution of the freezing rule to the overall root sublattice $L_\Lambda$ of $\Lambda$. 

\begin{table}
	\begin{center}
		\begin{tabular}{|c|c|c|c|c|}\hline
			Algebra & $k_\Lambda$ &Order of $k_\Lambda$& Transforms to & Contribution to $\Lambda$\\ \hline
			$\mra_{qn-1}$&$n$&$q = 2,3,4,5,6$&$\mra_{n-1}$ & $n \mra_{q-1}$ \\  \hline
			$\mrd_{n+2}$&$v$&$2$&$\mrc_{n}$& $2 \mra_{1}$\\ \hline
			$\mrd_{2n}$&$s$&$2$&$\mrb_n$& $n \mra_{1}$\\ \hline
			$\mre_7$&1&$2$&$\mrf_4$& $3 \mra_{1}$ \\ \hline
			$\mre_6$&1&$3$&$\mrg_2$& $2 \mra_{2}$\\\hline
            $\mrd_{2n+3}$&$1\simeq s$&$4$&$\mrc_n$& $n\mra_1+\mra_3$\\ \hline
   \end{tabular}
		\caption{Freezing rules for the simple factors in the gauge groups according to the element $k_\Lambda$ of the fundamental group associated to the freezing. For all the cases, the longest roots are of length twice the order of $k_\Lambda$. $v$ and $s$ denote the vector and spinor classes of the orthogonal groups.}
		\label{tab:rules}
	\end{center}
\end{table}

These transformations also appear in a seemingly unrelated problem, namely in the relation between the moduli space components of flat bundles over $T^2$ with non-simply-connected structure group $G$ \cite{Lerche:1997rr} when $G$ is simply-laced. The transformed group is simply-connected and describes the so-called topologically non-trivial components of the moduli space for a certain $G$. In this sense, what we find in the moduli space of 6d heterotic strings is a generalization to semisimple lie groups with many factors and more complicated fundamental groups\footnote{
We are not aware of a treatment of this general problem in the literature.}. 

\section{Summary of results and outlook}
\label{sec:results}

Let us summarize our results. The connected components of moduli space of the heterotic string studied in this paper have momentum lattices and corresponding orthogonal complements in $\Gamma_{4,20}$ (frozen sublattices) as shown in Table \ref{tab:lattices}. Here we have given $\Lambda$ in terms of its root sublattice $L_\Lambda$ and the fundamental group of the gauge group associated to $\Lambda$\footnote{The $L_\Lambda$'s correspond to the singularities of $\mathrm{K3}\times S^1$ orbifolds of order $n$ in the dual M-theory \cite{deBoer:2001wca}.}. The gauge symmetry groups that can be realized in the $n = 2,...,6$ components can be obtained by applying a set of ``freezing rules" to those of the $n = 1$ one. To check if one of these freezings can be done with a certain $G$, one looks for order $n$ elements $k_\Lambda$ in $\pi_1(G)$ such that they define an embedding of $L_\Lambda$ into the root lattice $L$ of $G$. If this is the case, one applies the rules shown in Table \ref{tab:rules} according to this embedding, and obtains the fundamental group of the resulting gauge group $G'$ by modding the elements of $\pi_1(G)$ by $k_\Lambda$ and projecting them onto the center of $G'$. Lists of gauge groups in these components can be found at \cite{fp2021}, and we give some examples of these freezings in Appendix \ref{app:examples}.

\begin{table}
	\begin{center}
		\begin{tabular}{|c|c|c|c|}\hline
			$n$ & Momentum Lattice $\Gamma$& Frozen root lattice $L_\Lambda$ & $\pi_1(G_\Lambda)$ \\ \hline
			$1$&$\Gamma_{4,20}$&$\emptyset$&\\  \hline
			$2$&$\Gamma_{3,3}\oplus \Gamma_{1,1}(2) \oplus \mrd_4 \oplus \mrd_4$&$8\mra_1$&$\mathbb{Z}_2$\\ \hline
			$3$&$\Gamma_{3,3}\oplus \Gamma_{1,1}(3) \oplus \mra_2 \oplus \mra_2$&$6\mra_2$&$\mathbb{Z}_3$\\ \hline
			$4$&$\Gamma_{3,3}\oplus \Gamma_{1,1}(4) \oplus \mra_1 \oplus \mra_1$&$2\mra_1 \oplus 4 \mra_3$ &$\mathbb{Z}_4$\\ \hline
			$5$&$\Gamma_{3,3}\oplus \Gamma_{1,1}(5)$&$4\mra_4$&$\mathbb{Z}_5$\\\hline
			$6$&$\Gamma_{3,3}\oplus \Gamma_{1,1}(6)$&$2\mra_1 \oplus 2\mra_2 \oplus 2\mra_5$&$\mathbb{Z}_6$\\ \hline
		\end{tabular}
		\caption{Momentum lattices and corresponding orthogonal complements in $\Gamma_{4,20}$, given in terms of their root sublattices and fundamental group of the associated gauge group.}
		\label{tab:lattices}
	\end{center} 
\end{table}

The moduli space components that we have studied are not all. In \cite{deBoer:2001wca} it was shown that there is a $\mathbb{Z}_2 \times \mathbb{Z}_2$-quadruple in 6d, but, in any case, an exhaustive list of the components of the moduli space of heterotic strings in 6d with maximal supersymmetry is not known. However, the map we have obtained is defined in terms of the fundamental group elements of the gauge groups and seems to naturally extend to many other cases that may correspond to other moduli space components, some of which require an M-theory description. This extension is the subject of an upcoming work.

On the other hand, the relation between these freezing rules and the problem of non-simply-connected flat $G$-bundles over $T^2$ is not clear, as in the heterotic string we are considering bundles over $T^4$. It may be better understood, perhaps, in a dual frame such as F-theory on $\mathrm{K}3\times T^2$ where one can more naturally isolate tori such as the fibers of the K3. As the former problem is rather high-level, it is tantalizing to think that it may play a role in constraining the possible theories with 16 supercharges that can be coupled to gravity (see e.g. \cite{Bedroya:2021fbu} for recent results in this direction).

\subsection*{Acknowledgements}
We are grateful to Mariana Graña, Ruben Minasian and Miguel Montero for stimulating conversations. This work was partially supported by the ERC Consolidator Grant 772408-Stringlandscape, PIP-CONICET-11220150100559CO, UBACyT 2018-2021 and ANPCyT-PICT-2016-1358 (2017-2020).
\newpage
\appendix

\section{Examples of gauge group freezings}
\label{app:examples}
Here we give some examples of freezings of gauge groups in 6d heterotic strings. For simplicity we use the A-to-G notation for gauge \textit{groups}. Whenever the length of an A factor is not 2, superscript indicates half its length. 
\vspace{1em}
\begin{enumerate}[label=\alph*)]
	
	\begin{minipage}{\textwidth}
		\item $2\mra_1 + \mra_{11} + \mrd_7$ with $H=\mathbb{Z}_{4}$ generated by $k=(0, 0, 3, 3)$, with center $(2,2,12,4)$ can be frozen to 
		\begin{center}
			\begin{tabular}{|c|c|c|c|c|c|}\hline 
				$k_{\Lambda}$ & Singularity  &		L & H & k & Center \\ \hline 
				 (0,0,6,2)&
     $8\mra_1/\mathbb{Z}_2$	&		$2\mra_1 + \mrc_5 + \mra_5^2$ & $\mathbb{Z}_2$ & (0,0,1,3) & (2,2,2,6) \\ \hline
				(0,0,9,1)& $(2\mra_1+4\mra_3)/\mathbb{Z}_{4}$	&	$2\mra_1 + \mrc_2^2 + \mra_2^4$ & 1 &	 & (2,2,2,3) \\ \hline
			\end{tabular}
		\end{center}
	\end{minipage}
	
	\vspace{.5em}\begin{minipage}{\textwidth}
		\item 
		$\mra_3 + \mra_{11} + \mre_6$ with $H=\mathbb{Z}_{6}$ generated by $k=(2,2,2)$, with center $(4,12,3)$ can be frozen to 
		\begin{center}
			\begin{tabular}{|c|c|c|c|c|c|}\hline 
				$k_{\Lambda}$ & Singularity  &		L & H & k & Center \\ \hline 
				(2,6,0)& $8\mra_1/\mathbb{Z}_2$	&	$\mre_6 + \mra_1^2 + \mra_5^2$ & $\mathbb{Z}_3$ &	(1,0,2) & (3,2,6) \\ \hline	
				(0,4,1)& $6\mra_2/\mathbb{Z}_3$	&	$\mra_3 + \mrg_2 + \mra_3^3$ & $\mathbb{Z}_2$ &	(2,0,2) & (4,1,4) \\ \hline
				(2,2,2)& $(2\mra_1+2\mra_2+2\mra_5)/\mathbb{Z}_{6}$	&	$\mrg_2 + \mra_1^2 + \mra_1^6$ & 1 &  & (1,2,2) \\ \hline
			\end{tabular}
		\end{center}
	\end{minipage}
	
	\vspace{.5em}\begin{minipage}{\textwidth}
		\item $3\mra_1 + 2\mra_4 + \mra_9$ with $H=\mathbb{Z}_{10}$ generated by $k=(1, 1, 1, 4, 4, 1)$, with center $(2,2,2,5,5,10)$ can be frozen to 
		\begin{center}
			\begin{tabular}{|c|c|c|c|c|c|}\hline 
				$k_{\Lambda}$ & Singularity  &		L & H & k & Center \\ \hline 
				(1,1,1,0,0,5)& $8\mra_1/\mathbb{Z}_2$	&	$2\mra_4 + \mra_4^2$ & $\mathbb{Z}_5 $ &	(4,4,1) & (5,5,5) \\ \hline
				(0,0,0,3,3,2)& $4\mra_4/\mathbb{Z}_5$	&	$3\mra_1 + \mra_1^5$ & $\mathbb{Z}_2$ & (1,1,1,1) & (2,2,2,2) \\ \hline	
			\end{tabular}
		\end{center}
	\end{minipage}
	
	\vspace{.5em}\begin{minipage}{\textwidth}
		\item $2\mra_2 + 2\mra_5 + \mre_6$ with $H=\mathbb{Z}_{3}^2$ generated by $k_1=(0,0,2,4,1)$ and $k_2=(1,2,0,2,1)$, with center $(3,3,6,6,3)$ can be frozen to 
		\begin{center}	
			\begin{tabular}{|c|c|c|c|c|c|}\hline 
				$k_{\Lambda}$ & Singularity  &		L & H & k & Center \\ \hline 
				(0,0,2,4,1)& $6\mra_2/\mathbb{Z}_3$	&	$2\mra_2 + \mrg_2 + 2\mra_1^3$ & $\mathbb{Z}_3$ & (1,1,0,0,0) & (3,3,1,2,2) \\ \hline
				(2,1,2,2,0)& $6\mra_2/\mathbb{Z}_3$	&	$\mre_6 + 2\mra_1^3$ & $\mathbb{Z}_3$ & (1,0,0) & (3,2,2) \\ \hline
				(1,2,0,2,1)& $6\mra_2/\mathbb{Z}_3$	&	$\mra_5 + \mrg_2 + \mra_1^3$ & $\mathbb{Z}_3$ & (2,0,0) & (6,1,2) \\ \hline
			\end{tabular}
		\end{center}
	\end{minipage}
	
	\vspace{.5em}\begin{minipage}{\textwidth}
		\item $3\mra_1 + \mrd_{10} + \mre_7$ with $H=\mathbb{Z}_{2}^2$ generated by $k_1=(0,0,0,1,0,1)$ and $k_2=(1,1,1,0,1,0)$, with center $(2,2,2,(2,2),2)$ can be frozen to 
		\begin{center}	
			\begin{tabular}{|c|c|c|c|c|c|}\hline 
				$k_{\Lambda}$ & Singularity  &		L & H & k & Center \\ \hline 
				(1,1,1,1,1,1)& $8\mra_1/\mathbb{Z}_2$	&	$\mrc_8 + \mrf_4$ & $\mathbb{Z}_2$ & (1,0) & (2,1) \\ \hline
				(1,1,1,0,1,0)& $8\mra_1/\mathbb{Z}_2$	&	$\mrb_5 + \mre_7$ & $\mathbb{Z}_2$ & (1,1) & (2,2) \\ \hline
				(0,0,0,1,0,1)& $8\mra_1/\mathbb{Z}_2$	&	$3\mra_1 + \mrb_5 + \mrf_4$ & $\mathbb{Z}_2$ & (1,1,1,1,0) & (2,2,2,2,1) \\ \hline
			\end{tabular}
		\end{center}
	\end{minipage}
	
	\vspace{.5em}\begin{minipage}{\textwidth}
		\item $2\mra_1 + 2\mra_2 + \mra_3 + \mra_{11}$ with $H=\mathbb{Z}_{12}$ generated by $k=(1,1,2,2,1,1)$, with center $(2,2,3,3,4,12)$ can be frozen to \vspace{-1.5em}\begin{center}
			\resizebox{\textwidth}{!}{%
				\begin{tabular}{|c|c|c|c|c|c|}\hline 
					$k_{\Lambda}$ & Singularity  &		L & H & k & Center \\ \hline 
					(0,0,0,0,2,6)& $8\mra_1/\mathbb{Z}_2$	&	$2\mra_1 + 2\mra_2 + \mra_1^2 + \mra_5^2$ & $\mathbb{Z}_6 $ &	(1,1,2,2,1,1) & (2,2,3,3,2,6) \\ \hline
					(0,0,2,2,0,4)& $6\mra_2/\mathbb{Z}_3$	&	$2\mra_1 + \mra_3 + \mra_3^3$ & $\mathbb{Z}_4 $ &	(1,1,1,1) & (2,2,4,4) \\ \hline	
					(1,1,0,0,3,3)& $(2\mra_1+4\mra_3)/\mathbb{Z}_4$ 	&	$2\mra_2  + \mra_2^4$ & $\mathbb{Z}_3$ & (2,2,1)   & (3,3,3)\\ \hline
					(0,0,1,1,2,2)& $(2\mra_1+2\mra_2+2\mra_5)/\mathbb{Z}_6$	&	$2\mra_1 + \mra_1^2 + \mra_1^6$ & $\mathbb{Z}_2$ & (1,1,1,1) & (2,2,2,2) \\ \hline
				\end{tabular}
		}\end{center}
	\end{minipage}
	
	\vspace{.5em}\begin{minipage}{\textwidth}
		\item $\mra_1 + 3\mra_5 + \mrd_4$ with $H=\mathbb{Z}_2 \times \mathbb{Z}_6$ generated by $k_1=(0,0,3,3,(1,1))$ and $k_2=(0,1,1,2,(0,1)))$, with center $(2,6,6,6,(2,2))$ can be frozen to 
		\vspace{-.5em}\begin{center}
			\resizebox{\textwidth}{!}{%
				\begin{tabular}{|c|c|c|c|c|c|}\hline 
					$k_{\Lambda}$ & Singularity  &		L & H & k & Center \\ \hline 
					(0,0,3,3,(1,1))& $8\mra_1/\mathbb{Z}_2$	&	$\mra_1 + \mra_5 + \mrc_2 + 2\mra_2^2$ & $\mathbb{Z}_6 $ &	(0,5,1,1,2) & (2,6,2,3,3) \\ \hline	
					(0,2,2,4,(0,0))& $6\mra_2/\mathbb{Z}_3$ 	&	$\mra_1 + \mrd_4 + 3\mra_1^3$ & $\mathbb{Z}_2^2$ &  $\begin{array}{c}
						(0,(1,0),0,1,1) \\
						(0,(0,1),1,0,1)
					\end{array}$  & (2,(2,2),2,2,2)\\ \hline
					(0,1,1,2,(0,1))& $(2\mra_1+2\mra_2+2\mra_5)/\mathbb{Z}_6$	&	$\mra_1 + \mrc_2 + \mra_1^3$ & $\mathbb{Z}_2$ & (0,1,1) & (2,2,2) \\ \hline	
				\end{tabular}
		}\end{center}
	\end{minipage}

	\vspace{.2em}	\vspace{.5em}\begin{minipage}{\textwidth}
		\item $4\mra_1 + \mra_2 + 2\mra_7$ with $H=\mathbb{Z}_2 \times \mathbb{Z}_4$ generated by $k_1=(0,0,1,1,0,2,2)$ and $k_2=(1,1,1,1,0,0,4)$, with center $(2,2,2,2,3,8,8)$ can be frozen to 
		\vspace{-1.5em}\begin{center}
			\resizebox{\textwidth}{!}{%
				\begin{tabular}{|c|c|c|c|c|c|}\hline 
					$k_{\Lambda}$ & Singularity  &		L & H & k & Center \\ \hline 
					(1,1,1,1,0,0,4)& $8\mra_1/\mathbb{Z}_2$ 	&	$\mra_2 + \mra_7 + \mra_3^2$ & $\mathbb{Z}_4$ & (0,2,2) & (3,8,4)\\ \hline
					(0,0,0,0,0,4,4)& $8\mra_1/\mathbb{Z}_2$	&	$4\mra_1 + \mra_2  + 2\mra_3^2$ & $\mathbb{Z}_2^2$ & $\begin{array}{c}
						(0,1,0,1,0,2,2) \\
						(1,0,1,0,0,2,2)
					\end{array}$ & (2,2,2,2,3,4,4) \\ \hline		
					(0,0,1,1,0,2,2)& $(2\mra_1+4\mra_3)/\mathbb{Z}_4$	&	$2\mra_1 + \mra_2 + 2\mra_1^4$ & $\mathbb{Z}_2$ & (1,1,0,0,0) & (2,2,3,2,2) \\ \hline
				\end{tabular}
		}\end{center}
	\end{minipage}
	
	\vspace{.5em}\begin{minipage}{\textwidth}
		\item $4\mra_1 + 2\mra_3 + 2\mrd_5$ with $H=\mathbb{Z}_{2}\times \mathbb{Z}_4$ generated by $k_1=(0, 0, 0, 0, 1, 3, 1, 3)$ and $k_2=(1,1,1,1,0,0,2,2)$, with center $(2,2,2,2,4,4,4,4)$ can be frozen to 	\vspace{-1.5em}\begin{center}
			\resizebox{\textwidth}{!}{%
				\begin{tabular}{|c|c|c|c|c|c|}\hline 
					$k_{\Lambda}$ & Singularity  &		L & H & k & Center \\ \hline 
					(1,1,1,1,2,2,0,0)& $8\mra_1/\mathbb{Z}_2$	&	$2\mrd_5 + 2\mra_1^2$ & $\mathbb{Z}_4 $ &	(1,3,1,1) & (4,4,2,2) \\ \hline	
					(1,1,1,1,0,0,2,2)& $8\mra_1/\mathbb{Z}_2$	&	$2\mra_3 + 2\mrc_3$ & $\mathbb{Z}_4 $ &	(1,3,1,1) & (4,4,2,2) \\ \hline	
					(0,0,0,0,2,2,2,2)& $8\mra_1/\mathbb{Z}_2$	&	$4\mra_1 + 2\mrc_3 + 2\mra_1^2$ & $\mathbb{Z}_2^2 $ &	$\begin{array}{c}
						(1,1,1,1,0,0,0,0) \\
						(0,0,0,0,1,1,1,1) \\
					\end{array}$ & (2,2,2,2,2,2,2,2) \\ \hline
					(0,0,0,0,1,3,1,3)& $(2\mra_1+4\mra_3)/\mathbb{Z}_4$	&	$4\mra_1 + 2\mra_1^4$ & $\mathbb{Z}_2$ &   $\begin{array}{c}
						(1,1,1,1,0,0) \\
					\end{array}$ & (2,2,2,2,2,2) \\ \hline	
				\end{tabular}
		}\end{center}
	\end{minipage}

	\vspace{.5em}\begin{minipage}{\textwidth}
		\item $5\mra_1 + \mrd_4 + \mrd_5 + \mrd_6$ with $H=\mathbb{Z}_{2}^3$ generated by $k_1=(0, 0, 0, 0, 1, (0,1), 2,(0,1))$, $k_2=(0,1,1,1,0,(0,0),2,(1,0)))$ and $k_3=(1,0,0,1,0,(1,1),2,(1,1)))$, with center $(2,2,2,2,2,(2,2),4,(2,2))$ can be frozen to
		\vspace{-1.5em}\begin{center}
			\resizebox{\textwidth}{!}{%
				\begin{tabular}{|c|c|c|c|c|c|}\hline 
					$k_{\Lambda}$ & Singularity  &		L & H & k & Center \\ \hline 
					(0,0,0,0,1,(0,1),2,(0,1))& $8\mra_1/\mathbb{Z}_2$	&	$4\mra_1+\mrb_3 + \mrc_2 + \mrc_3$ & $\mathbb{Z}_2^2$ &   $\begin{array}{c}
						(0,1,1,1,0,1,0) \\
						(1,0,0,1,1,1,0) 
					\end{array}$ &  (2,2,2,2,2,2) \\ \hline
					(0,1,1,1,0,(0,0),2,(1,0))& $8\mra_1/\mathbb{Z}_2$	&	$2\mra_1+\mrb_3 + \mrc_3 + \mrd_4$ & $\mathbb{Z}_2^2$ &   $\begin{array}{c}
						(0,1,1,0,(0,1)) \\
						(1,0,1,0,(1,1)) 
					\end{array}$ &  (2,2,2,2,(2,2)) \\ \hline
					(0,1,1,1,1,(0,1),0,(1,1))& $8\mra_1/\mathbb{Z}_2$	&	$\mra_1+\mrc_2 + \mrc_4 + \mrd_5$ & $\mathbb{Z}_2^2$ &   $\begin{array}{c}
						(0,0,1,2) \\
						(1,1,0,2) 
					\end{array}$ &  (2,2,2,4) \\ \hline
					(1,0,0,1,0,(1,1),2,(1,1))& $8\mra_1/\mathbb{Z}_2$	&	$3\mra_1+\mrc_2 + \mrc_3 + \mrc_4$ & $\mathbb{Z}_2^2$ &   $\begin{array}{c}
						(0,1,1,0,0,1) \\
						(1,0,0,1,0,1) 
					\end{array}$ &  (2,2,2,2,2,2) \\ \hline	
					(1,0,0,1,1,(1,0),0,(1,0))& $8\mra_1/\mathbb{Z}_2$	&	$2\mra_1+\mrb_3+\mrc_2 + \mrd_5$ & $\mathbb{Z}_2^2$ &   $\begin{array}{c}
						(0,0,1,1,2) \\
						(1,1,0,0,2) 
					\end{array}$ &  (2,2,2,2,4) \\ \hline	
					(1,1,1,0,0,(1,1),0,(0,1))& $8\mra_1/\mathbb{Z}_2$	&	$2\mra_1+\mrb_3+\mrc_2 + \mrd_5$ & $\mathbb{Z}_2^2$ &   $\begin{array}{c}
						(1,0,1,0,2) \\
						(0,1,0,1,2) 
					\end{array}$ &  (2,2,2,2,4) \\ \hline	
					(1,1,1,0,1,(1,0),2,(0,0))& $8\mra_1/\mathbb{Z}_2$	&	$\mra_1+\mrc_2+\mrc_3+\mrd_6$ & $\mathbb{Z}_2^2$ &   $\begin{array}{c}
						(0,1,0,(1,0)) \\
						(1,0,0,(0,1)) 
					\end{array}$ &  (2,2,2,(2,2)) \\ \hline	
				\end{tabular}
		}\end{center}
	\end{minipage}

\end{enumerate}
\bibliographystyle{JHEP}
\bibliography{Heterotic6d}

\end{document}